\documentclass[onecolumn,tightenlines,showpacs,nofootinbib,preprintnumbers,
amsmath,amssymb]{revtex4}

\usepackage[usenames,dvipsnames]{color}

\usepackage{graphicx}
\usepackage{bm}
\usepackage{lscape}

\usepackage{setspace}
\begin{document}
\begin{spacing}{1.5}


\title{Detection of supernova neutrinos on the Earth for large $\theta_{13}$}

\author{Jing Xu$^{1}$\footnote{Email: 200921220002@mail.bnu.edu.cn}, Ming-Yang Huang $^{2}$\footnote{Email: huangmy@ihep.ac.cn}, Xin-Heng Guo$^{1}$\footnote{Corresponding author, Email: xhguo@bnu.edu.cn} and Bing-Lin Young$^{3,4}$\footnote{Email: young@iastate.edu}}
\affiliation{\small $^{1}$College of Nuclear Science and
Technology, Beijing Normal University, Beijing 100875, China\\
\small $^{2}$Institute of High Energy Physics, Chinese Academy of
Sciences, Beijing 100049, China\\
\small $^{3}$Department of Physics and Astronomy, Iowa State
University, Ames, Iowa 5001, USA\\
$^{4}$Institute of Theoretical Physics, Chinese Academy of Sciences, Beijing 100190, China }

\begin{abstract}
Supernova (SN) neutrinos detected on the Earth are subject to the shock wave effects, the Mikheyev-Smirnov-Wolfenstein (MSW) effects, the neutrino collective effects and the Earth matter effects. Considering the recent experimental result about the large mixing angle $\theta_{13}$ ($\backsimeq8.8^{\circ}$) provided by the Daya Bay Collaboration and applying the available knowledge for the neutrino conversion probability in the high resonance of SN, $P_{H}$, which is in the form of hypergeometric function in the case of large $\theta_{13}$, we deduce the expression of $P_{H}$ taking into account the shock wave effects. It is found that $P_{H}$ is not zero in a certain range of time due to the shock wave effects. After considering all the four physical effects and scanning relevant parameters, we calculate the event numbers of SN neutrinos detected at the Daya Bay experiment. From the numerical results, it is found that the behaviors of neutrino event numbers detected on the Earth depend on the neutrino mass hierarchy and neutrino spectrum parameters including the temperature $T_{\alpha}$, the dimensionless pinching parameter $\eta_{\alpha}$ or $\beta_{\alpha}$ (where $\alpha$ refers to neutrino flavor), the average energy $\langle E_{\alpha}\rangle$, and the SN neutrino luminosities $L_{\alpha}$. We also compare the results of two parametrization methods for the neutrino energy distributions and give the ranges of SN neutrino event numbers that will be detected at the Daya Bay experiment.

\end{abstract}


\maketitle

\section{Introduction}

In the universe, core-collapse supernovas (SNs) are among the most energetic explosions \cite{Kotake}-\cite{Bethe1}. They not only mark the catastrophic end of some stars, which turn into neutron stars or black holes after explosions, but also are responsible for the richness of heavy elements \cite{Lattimer}-\cite{Lal}. SN1987A has attracted worldwide interests and has been studied extensively since it came into our sight several decades ago \cite{Arnett1}\cite{Bionta}. During the explosions of the type II SN, most of the binding energy is released as neutrinos, which are very useful for acquiring information about intrinsic properties and the explosion dynamics of SN \cite{Bethe1}\cite{Arnett1}.

For the past few decades, in most theoretical models, it has been believed that the neutrino mixing angle $\theta_{13}$ is smaller than $3^{\circ}$, or even smaller than $1.5^{\circ}$. In Refs. \cite{Guo}\cite{Huang}, the authors studied possible methods to measure this neutrino mixing angle while $\theta_{13}<3^{\circ}$. However, in recent years, some new experimental results indicated a large $\theta_{13}$ (by large $\theta_{13}$ we mean $\theta_{13}\backsimeq9^{\circ}$) \cite{Ardellier}-\cite{Abe2}. This year the Daya Bay experiment measured the value of $\theta_{13}$ to $5.2\sigma$ accuracy and obtained the result $\theta_{13}=8.8^{\circ}\pm0.8^{\circ}$ \cite{An1}\cite{Daya}, which is much larger than $3^{\circ}$. Table I shows a summary of the recent experimental results about $\theta_{13}$. In this paper, we will study detection of SN neutrinos on the Earth in the case of large $\theta_{13}$.

\begin{table}[!htb]
\begin{center}
\caption{Summary of experimental results about $\theta_{13}$. Numbers with (without) brackets are for normal (inverted) mass hierarchy.}
\vspace{0.3cm}
\begin{tabular}{|c|c|c|c|c|}\hline
Experiment   & Result & Accuracy(Sensitivity)	& Year & Reference \\
\hline
Double Chooz & $0.03<\sin^22\theta_{13}<0.19$  & $0.03$   & 2006 &\cite{Ardellier} \\
\hline
MINOS        & $0<\sin^22\theta_{13}<0.12(0.19)$   & $1\sigma$  & 2009 & \cite{Wang}\\
\hline
T2K          & $0.03(0.04)<\sin^22\theta_{13}<0.28(0.34)$   & $2.5\sigma$ & 2010 & \cite{Abe1} \\
\hline
MINOS        & $2\sin^22\theta_{23}\sin^22\theta_{13}<0.12(0.20)$   & $90\%$   & 2011 & \cite{Adamson} \\
\hline
Daya Bay      & $\sin^22\theta_{13}=0.092\pm0.016(stat)\pm0.005(syst)$   & $5.2\sigma$ & 2012 & \cite{An1}\\
\hline
RENO & $\sin^22\theta_{13}=0.086\pm0.041(stat)\pm0.03(syst)$  & $4.9\sigma$  & 2012 & \cite{Abe2} \\
\hline
\end{tabular}
\end{center}
\end{table}

SN neutrinos are produced from the core collapse of SN and propagate outward to the surface of SN. Then they travel a long cosmic distance to reach the detector on the Earth. During this process, they pass through the SN matter and the Earth matter. While they propagate, SN neutrinos are subject to the Mikheyev-Smirnov-Wolfenstein (MSW) effects \cite{Wolfenstein1}\cite{Kuo}, the shock wave effects \cite{Takiwaki}\cite{Fogli1}, the neutrino collective effects \cite{Dasgupta}, and the Earth matter effects \cite{Guo}\cite{Huang}\cite{Lunardini1}. Different from the case in vacuum, the behavior of neutrino oscillation changes while neutrinos propagate in matter. The neutrino matter effects, due to the interaction between matter and neutrinos, was found by Wolfenstein, Mikheyev and Smirnov, and was named as MSW effects. Inside the SN, the large mixing angle solution of neutrinos results in the crossing probability at the low resonance region $P_{L}\sim0$. Therefore, we only need to consider the crossing possibility at the high resonance region, $P_{H}$ \cite{Kuo}\cite{Mikheyev}. In addition, the shock wave effects may change the density distribution of SN and the position of the high resonance. The expression of $P_{H}$ for large $\theta_{13}$ needs to be developed \cite{Huang}\cite{Kuo}\cite{Kachelriess}. When studying the Earth matter effects, the realistic density distribution of the Earth needs to be considered \cite{Dziewonski}\cite{Stacey}. By scanning the ranges of relevant parameters appearing in the neutrino spectra, it is possible to obtain the maximum and minimum event numbers of SN neutrinos detected on the Earth. There are two different parametrization forms for the neutrino energy distribution \cite{Kotake}\cite{Chakraborty}. We will compare the results from these two models.

This paper is organized as follows. In Section II, we present a very brief overview of SN explosions and the production of SN neutrinos. In Section III, the four physical effects on the detection of SN neutrinos including the MSW effects, the shock wave effects, the neutrino collective effects, and the Earth matter effects are discussed respectively. In this section, the expression of $P_{H}$ in the high resonance region for large $\theta_{13}$ is obtained. In Section IV, using the latest result of the Daya Bay experiment, $\theta_{13}= 8.8^{\circ}\pm0.8^{\circ}$, we take into account the four physical effects and calculate the event numbers of SN neutrinos detected on the Earth. The results of two different parametrization methods for the neutrino energy distribution are compared and the effects of relevant SN neutrino parameters are discussed in detail. Finally, in Section V we give the summary and discussions.

\section{SN explosions and SN neutrino spectra}

According to the presence or absence of hydrogen lines in their spectra, supernovas can be classified into two types, type I and type II. The so-called type II supernovas have hydrogen lines in their spectra, and are core-collapse supernovas. In this paper, we only pay attention to the type II supernovas, which are one main source of neutrinos in the universe \cite{Loredo1}. The explosion process of core-collapse supernova can be divided into several phases, and more details about the scenario of explosion can be found in Ref. \cite{Kotake}.

An SN explosion approximately releases a total energy of $E_B=3\times10^{53}erg$, about $99\%$ of which is radiated away as SN neutrinos \cite{Spergel}. The relation between the total SN energy and the luminosity of different flavor neutrinos is given by \cite{Fogli0}
\begin{equation}
L_{{\nu}_{e}}(t)+ L_{\bar{{\nu}}_{e}}(t)+ L_{{\nu}_{x}}(t)=\frac{E_{B}}{\tau}e^{-t/\tau}.
\end{equation}
where ${\nu}_x$ represent ${\nu}_{\mu}$, ${\nu}_{\tau}$, ${\bar{\nu}}_{\mu}$ and ${\bar{\nu}}_{\tau}$. The luminosity flux of the SN neutrinos $L_{\alpha}$ ($\alpha={\nu}_{e},{{\bar{\nu}}_{e}},{{\nu}_{x}}$) decays in time as \cite{Spergel}\cite{Loredo2}\cite{Totsuka}
\begin{equation}
L_{\alpha}(t)=L^0_{\alpha}e^{-t/\tau}. \label{L}
\end{equation}
The range of $\tau$ was obtained by fitting the experimental data of SN1987A: $\tau=1.74-4.19s$ \cite{Spergel}\cite{Loredo2}.

In general, the SN neutrino spectra are parameterized in two forms to match the result of Monte Carlo simulations. One is known as the Fermi-Dirac distribution \cite{Lunardini2},
\begin{equation}
F_{\alpha}^{(0)}(E)=\frac{L_{\alpha}}{F_{\eta_{\alpha}}T_{\alpha}^{4}}
\frac{E^2}{e^{(E/T_{\alpha}-\eta_\alpha)}+1},
\label{Foa}
\end{equation}
where $F_{\eta_{\alpha}}$ is defined as:
\begin{equation}
F_{\eta_{\alpha}}=\int_{0}^{\infty}\frac{x^3}{\exp{(x-\eta_\alpha)}+1}{\rm
d}x \nonumber,
\end{equation}
and $E, \eta_\alpha, T_{\alpha}$ are the neutrino energy, the pinching parameter of the spectrum, and the neutrino temperature, respectively.
The spectra obtained from numerical simulations can be well fitted by choosing \cite{Janka}
\begin{eqnarray}
T_{{\nu}_{e}}=3-4MeV, & T_{\bar{\nu}_{e}}=5-6MeV, & T_{{\nu}_{x}}=7-9MeV,
\label{T}\\
\eta_{{\nu}_{e}}\approx3-5, & \eta_{\bar{\nu}_{e}}\approx2.0-2.5, &
\eta_{{\nu}_{x}}\approx0-2. \nonumber
\end{eqnarray}
In addition, the luminosity ratios of different flavor neutrinos, which will be shown later to play an important role in the calculation of event numbers, are as follows:
\begin{eqnarray}
\frac{L_{{\nu}_{e}}}{L_{{\nu}_{x}}}=0.5-2,   & & \frac{L_{\bar{\nu}_{e}}}{L_{{\nu}_{x}}}=0.5-2. \label{L1}
\end{eqnarray}

The other parametrization form of SN neutrino fluxes given by the Garching group can be expressed as
\begin{equation}
F_{\alpha}^{(0)}(E)=\frac{L_{\alpha}}{\langle E_{\alpha}\rangle}\frac{\beta^{\beta_{\alpha}}_{\alpha}}{\Gamma(\beta_{\alpha})}
\Bigg(\frac{E}{\langle E_{\alpha}\rangle}\Bigg)^{(\beta_{\alpha}-1)}exp\Bigg(-\beta_{\alpha}\frac{{E}}{\langle E_{\alpha}\rangle}\Bigg),
\label{Fia}
\end{equation}
where $\langle E_{\alpha}\rangle$ is the average energy of neutrino and $\beta_{\alpha}$ is the dimensionless pinching parameter. For different neutrinos, their ranges of values are obtained as
\cite{Keil}
\begin{eqnarray}
\langle E_{{\nu}_{e}}\rangle=12-15MeV, & \langle E_{\bar{\nu}_{e}}\rangle=12-15MeV,
& \langle E_{{\nu}_{x}}\rangle=15-18MeV,
\label{E}\\ &
\beta_{\alpha}=3.5-6.
\nonumber
\end{eqnarray}
The ranges of luminosity ratios for different flavor neutrinos are the following in this model:
\begin{eqnarray}
\frac{L_{{\nu}_{e}}}{L_{{\nu}_{x}}}=0.5-0.8, & &   \frac{L_{\bar{\nu}_{e}}}{L_{{\nu}_{x}}}=0.5-0.8. \label{L2}
\nonumber\\
\end{eqnarray}

\section{Physical effects on the Detection of SN neutrinos}

When neutrinos propagate outward to the surface of SN, they can be subject to the SN shock wave effects, the MSW effects, and the collective effects. Before arriving at the detectors, they travel through the Earth matter and are affected by the Earth matter effects. In this section, we shall consider all the above four physical effects on SN neutrinos.

\subsection{MSW effects and conversion probability}

The MSW effects are caused by neutrino interactions with matter, which are determined by the matter density profile and the mixing angles. By using the Landau's method, the conversion probability $P_H$ for neutrinos to jump from one mass eigenstate to another at the high resonance layer can be expressed as \cite{Kuo}
\begin{eqnarray}
P_H=\frac {\exp{[{-\pi}{\gamma}{F}/{2}]}-\exp[{-\pi}{\gamma}{F}/{2{\sin^2}{\theta}}]}
{1-\exp{[{-\pi}{\gamma}{F}/{2{\sin^2}{\theta}}]}}. \label{PH}
\end{eqnarray}
The factor $F$ is given by
\begin{equation}
F=\begin{cases}1 & (n_e\propto r) \\
{(1-{\tan}^2{\theta})^2}/{(1+{\tan}^2{\theta})^2} & (n_e\propto r^{-1}) \\
(1-{\tan}^2{\theta}) & (n_e\propto e^{-r}) \\
2 \sum\limits_{m=0}^{\infty}\Bigg(
\begin{array}{cc}
1/n-1\\
2m
\end{array}
\Bigg)\Bigg[
\begin{array}{cc}
\frac{1}{2}\\
m+1
\end{array}
\Bigg]{({\tan}2{\theta})^{2m}} & (n_e\propto r^{n}), \\
\end{cases}
\label{F0}
\end{equation}
where $n_e$ is the electron density, $r$ is the distance to the center of SN, and the adiabaticity parameter $\gamma$ is defined as \cite{Kuo}
\begin{equation}
\gamma \equiv \frac{{\Delta} m^2}{E|{{\partial}{\ln} {n_e}}/{{\partial} r}|}{{\sin 2{\theta}}{\tan 2{\theta}}}, \label{gamma}
\end{equation}
with ${\Delta}m^2$ being the mass square difference of two mass eigenstates.

For the SNs, $n\approx-3$, the expression of $F$ in the case of $n_e\propto r^{n} $ in Eq. (\ref{F0}) is
\begin{equation}
F= 2 \sum\limits_{m=0}^{\infty}\Bigg(
\begin{array}{cc}
1/n-1\\
2m
\end{array}
\Bigg)\Bigg[
\begin{array}{cc}
\frac{1}{2}\\
m+1
\end{array}
\Bigg]{({\tan}2{\theta})^{2m}}.
\label{F1}
\end{equation}
In Eq. (\ref{F1})
\begin{equation}
\Bigg( \begin{array}{cc}
1/n-1\\
2m \end{array}
\Bigg)
=\frac{(1/n-1)!}{(1/n-1-2m)!(2m)!},
\end{equation}
and
\begin{equation}
2\Bigg[
\begin{array}{cc}
\frac{1}{2}\\
m+1
\end{array}
\Bigg]=(-1)^m \frac{J_m-J_{m+1}}{{\pi}/4},
\end{equation}
with
\begin{equation}
J_m=\int_{0}^{{\pi}/2}(\sin{\phi})^{2m}d{\phi}=\frac{(2m-1)!!}{(2m)!!}\frac{\pi}{2}
\end{equation}
Eq. (\ref{F1}) can be expressed as a hypergeometric function:
\begin{equation}
F= {_2}F_1\Bigg(\frac{n-1}{2n},\frac{2n-1}{2n};2;{{-\tan}^22\theta}\Bigg).
\end{equation}
In the case of $\theta\in[0, {\pi}/8]$, using the Euler integral representation \cite{Prudnikov}, one has
\begin{equation}
{_2}F_1(a,b;c;z)
=\frac{\Gamma(c)}{\Gamma(b)\Gamma(c-b)}\int_{0}^{1}t^{b-1}(1-t)^{c-b-1}(1-tz)^{-a}{\rm
d}t.
\label{F21}
\end{equation}
We make the Taylor expansion for $F$ near the point $\frac{1}{n}=0$,
\begin{equation}
F=F(0)+F^{'}(0)\Bigg(\frac{1}{n}\Bigg)+F^{''}(0)\Bigg(\frac{1}{n}\Bigg)^{2}+\cdots+F^{(m)}(0)\Bigg(\frac{1}{n}\Bigg)^{m}+\cdots.
\label{FF}
\end{equation}
The first two coefficients in Eq. (\ref{FF}) can be obtained straightforwardly,
\begin{eqnarray}
F(0)=1-\tan^{2}{\theta}, & &
F^{'}(0)=(1-\tan^{2}{\theta})\Bigg[\ln{(1-\tan^{2}{\theta})}+1
-\frac{1+\tan^{2}{\theta}}{\tan^{2}{\theta}}\ln(1+\tan^2{\theta})\Bigg].
\end{eqnarray}
Comparing with the numerical result of the right-hand side of Eq. (\ref{F21}) in the case $\frac{1}{n}\rightarrow0$, we find that the first two terms in Eq. (\ref{FF}) give dominant contributions and other items are negligible, so $F$ can be approximately written as
\begin{equation}
F=(1-\tan^2{\theta})
\Bigg(1-\frac{1}{n}\{{\ln(1-\tan^2{\theta})+1-[(1+\tan^2{\theta})/{\tan^2{\theta}}]}
{\ln(1+\tan^2{\theta})}\}\Bigg).
\label{FT}
\end{equation}
This expression is identical to that in \cite{Kuo}.

For the case of $n=-3$, the comparison between the numerical result of the right-hand side of Eq. (\ref{F21}) and that given by Eq. (\ref{FT}) is shown in Fig. 1. It can be seen that for $n=-3$, Eq. (\ref{FT}) is a very good approximation to $_2F_1$ in Eq. (\ref{F21}).

\subsection{SN shock wave effects and $P_H$}

The SN shock wave effects play an important role in the SN neutrino oscillations. As was pointed out in Ref. \cite{Schirato} and further studied in Refs. \cite{Takahashi1}-\cite{Fogli2}, after the core bounce, the shock wave propagates inside the SN during the period of neutrino emission. It modifies the density profile of the star and the change is characterized by a density jump as shown in Fig. 2 \cite{Huang}\cite{Fogli2}. In several seconds, the shock wave may reach the resonance region where the conversion of different flavor SN neutrinos maximize, thus affecting the transition probability $P_H$ in the high resonance region \cite{Takiwaki}\cite{Mikheyev}. The density distribution of SN might be divided into two phases roughly by time. Before the shock wave effects begin, i.e. $t<1s$, the SN matter density is \cite{Schirato}:
\begin{equation}
 \rho_0(r)\backsimeq 10^{14}\cdot\Bigg(\frac{r}{1km}\Bigg)^{-2.4} g/cm^3.
 \label{rho0}
\end{equation}
In Eq. (\ref{rho0}) $n=-2.4$. We have checked that in this case Eq. (\ref{FT}) is a very good approximation to Eq. (\ref{F21}). The difference between the numerical results from these two equations are negligible when $\theta_{13}=8.8^{\circ}$.

For $t\geq1s$, the shock wave effects set in and the matter density is given by \begin{equation}
\rho(r,t)=\rho_0(r)\cdot
\begin{cases}\xi\cdot f(r,t) & (r\leqslant r_s), \\
 1 & (r>r_s),
 \end{cases}
\label{rho}
\end{equation}
where $r_s$ is the position of the shock wave front, $f(r,t)$ is defined as \cite{Schirato}
\begin{equation}
 f(r,t)=\exp\{[0.28-0.69\ln(r_s/km)][\arcsin(1-r/r_s)]^{1.1}\},
 \label{f}
\end{equation}
and $\xi$ is a typical ratio of the potential across the shock wave
front,
\begin{eqnarray}
 \xi=V_+/V_-\backsimeq10, \label{xi}
\end{eqnarray}
which measures the SN matter potential $V(r)$ drop from
\begin{eqnarray}
V_+=\lim_{r\rightarrow r_s^-}V(r), \label{V+}
\end{eqnarray}
to
\begin{eqnarray}
V_-=\lim_{r\rightarrow r_s^+}V(r). \label{V-}
\end{eqnarray}
The SN matter potential is related to the SN electron density
$n_e(r)$ by
\begin{eqnarray}
V(r)=\sqrt{2}G_F n_e(r)=\sqrt{2}G_F N_A\rho(r) Y_e \label{VS},
\end{eqnarray}
where $G_F$ is the Fermi constant, $N_A$ is the Avogadro's number and $Y_e$ is the electron fraction. In the numerical calculations, we assume $Y_e=0.5$. In Eq. (22) and Eq. (23), a slightly accelerating shock-wave-front position $r_s$ is assumed with the explicit time dependence \cite{Schirato},
\begin{eqnarray}
r_s(t)=-4.6\times10^{3}+1.13\times10^{4}\cdot t+1\times10^{2}\cdot t^2,
\label{rs}
\end{eqnarray}
where $r_s$ is in units of $km$ and $t$ in units of $s$. As shown in Fig. 2, there is a jump in the density curves when $t\geq1s$.

Suppose on density curves there are three points ($r=r_1, r_2, r_3$) satisfying the resonance condition, that is
\begin{eqnarray}
\Delta m^2_{31}\cdot \cos2\theta_{13}=2\sqrt{2}G_Fn_e(r_i)E\quad
(i=1,2,3), \label{resonance}
\end{eqnarray}
where $|\Delta
m^2_{31}|\backsimeq|\Delta m^2_{32}|\backsimeq2.4\times10^{-3} eV^2$
\cite{Gonzalez-Garcia}.
In this case, two of the points lie below the shock front and the third one beyond the shock front, i.e., $r_1 < r_2 < r_s < r_3 $, four crossing probabilities corresponding to these points are denoted by $P_{H1}$, $P_{H2}$, $P_{s}$ and $P_{H3}$, respectively. We also define four densities: $\rho_{res}$ is the resonance density, which satisfies Eq. (\ref{resonance}) \cite{Kotake}; $\rho_+$ is the density at $r=r_s$ corresponding to the matter potential $V_+$ and $\rho_-$ is the density at $r=r_s$ corresponding to the matter potential $V_-$; $\rho_b$ is the density at the bottom of the camber which is the minimum of the density profile below the shock front. Then in the case where three points satisfy the resonance condition, $\rho_b<\rho_{res}\leq\rho_-$, and $P_H$ has the most complicated expression:
\begin{eqnarray}
P_H&=&(P_{H1}+P_{H2}+P_{s}+P_{H3})-2(P_{H1}P_{H2}+P_{H1}P_{H3}+P_{H1}P_{s}
+P_{H2}P_{H3}+P_{H2}P_{s}+P_{H3}P_{s}) \nonumber\\
&&
+4(P_{H1}P_{H2}P_{H3}+P_{H1}P_{H2}P_{s}+P_{H1}P_{H3}P_{s}+P_{H2}P_{H3}P_{s})-8P_{H1}P_{H2}P_{H3}P_s
\nonumber\\
&&+2(1-2P_s-2P_{H3}+4P_{H3}P_s)\sqrt{P_{H1}P_{H2}(1-P_{H1})(1-P_{H2})}\cos\phi_{12}
\nonumber\\
&&+2(1-2P_{H1}-2P_{H3}+4P_{H1}P_{H3})\sqrt{P_{H2}P_{s}(1-P_{H2})(1-P_{s})}\cos\phi_{2s}
\nonumber\\
&&+2(1-2P_{H2}-2P_{H3}+4P_{H2}P_{H3})\sqrt{P_{H1}P_{s}(1-P_{H1})(1-P_{s})}\cos\phi_{1s}
\label{PH5}\\
&&+2(1-2P_{H1}-2P_{H2}+4P_{H1}P_{H2})\sqrt{P_{H3}P_{s}(1-P_{H3})(1-P_{s})}\cos\phi_{s3}
\nonumber\\
&&+2(1-2P_s-2P_{H1}+4P_{H1}P_s)\sqrt{P_{H2}P_{H3}(1-P_{H2})(1-P_{H3})}\cos\phi_{23}
\nonumber\\
&&+2(1-2P_s-2P_{H2}+4P_{H2}P_s)\sqrt{P_{H1}P_{H3}(1-P_{H1})(1-P_{H3})}\cos\phi_{13}
\nonumber\\
&&-8\sqrt{P_{H1}P_{H2}(1-P_{H1})(1-P_{H2})}\cos\phi_{12}\sqrt{P_{H3}
P_{s}(1-P_{H3})(1-P_{s})}\cos\phi_{s3},  \nonumber
\end{eqnarray}
where the crossing probability $P_{Hi}$ at $r_i$ ($i=1, 2, 3$) can be calculated from Eq. (9), $\phi_{ij}$ ($i,j=1, 2, 3, s$) is defined as
\begin{eqnarray}
\phi_{ij}\approx\int^{r_j}_{r_i}{\rm
d}x\frac{1}{2E}\sqrt{[\Delta{m}^2_{31}
   \cos2\theta_{13}-2EV(r)]^2+(\Delta{m}^2_{31} \sin2\theta_{13})^2}.
\label{phiij}
\end{eqnarray}

The number of the points where the resonance condition occurs might be two, one, or even none. In the case where $\rho_b<\rho_{res}\leq\rho_+$ and $\rho_{res}>\rho_-$, or $\rho_{res}=\rho_b$ and $\rho_{res}\leq\rho_-$, the resonance condition occurs at two points, $r=r_1, r_2$ or $r=r_1, r_3$, then
\begin{eqnarray}
P_H&=&(P_{H1}+P_{Hl}+P_{s})-2(P_{H1}P_{Hl}+P_{H1}P_{s}+P_{Hl}P_{s})+4P_{H1}P_{Hl}P_{s}
\nonumber\\
&&+2(1-2P_s)\sqrt{P_{H1}P_{Hl}(1-P_{H1})(1-P_{Hl})}\cos\phi_{1l}
\label{PH4}\\
&&+2(1-2P_{H1})\sqrt{P_{Hl}P_{s}(1-P_{Hl})(1-P_{s})}\cos\phi_{ls}
\nonumber\\
&&+2(1-2P_{Hl})\sqrt{P_{H1}P_{s}(1-P_{H1})(1-P_{s})}\cos\phi_{1s}
\quad (l=2, 3).   \nonumber
\end{eqnarray}

When $\rho_{res}>\rho_+$, or $\rho_{res}=\rho_b$ and
$\rho_{res}>\rho_-$, or $\rho_{res}<\rho_b$ and
$\rho_{res}\leq\rho_-$, the resonance condition occurs at only one point, $r=r_1$ or $r=r_3$, one obtains
\begin{eqnarray}
P_H=P_{Hk}+P_{s}-2P_{Hk}P_{s}+2\sqrt{P_{Hk}P_{s}(1-P_{Hk})(1-P_{s})}
\cos\phi_{ks}, \quad (k=1, 3). \label{PH3}
\end{eqnarray}

If $\rho_-<\rho_{res}<\rho_b$, the resonance condition does not occur, it can be easily realized that
\begin{eqnarray}
P_H=P_s.   \label{PH2}
\end{eqnarray}

A summary for neutrino flavor conversions due to the neutrino shock wave effects and the MSW effects in various density regions is given in Table II \cite{Huang}.

\begin{table}[hbtp!]
\begin{center}
\caption{Summary of neutrino flavor conversions due to the neutrino shock wave effects and the MSW effects in various density regions.}
\label{tab:Probab} \vspace{0.3cm}
\begin{tabular}{|c||c|c||c|c|}\hline
$t<1s$ & \multicolumn{4}{c|}{$t\geq 1s$} \\ \hline\hline
    & \multicolumn{2}{c||}{$\rho_b>\rho_-$}
     & \multicolumn{2}{c|}{$\rho_-\geq\rho_b$}
                \\ \cline{2-5}  
  & resonance in & flavor conversion
     & Resonance in & flavor conversion  \\
  & region & involved & region  & involved
                 \\ \cline{2-5}
  $P_{H3}$ & $\rho_{res}>\rho_+$ & $P_{H1},~P_s$
     & $\rho_{res}>\rho_+$ & $P_{H1},~P_s$
                 \\ \cline{2-5} 
  & $\rho_+\geq\rho_{res}>\rho_b$ & $P_{H1},~P_{H2},~P_s$
       & $\rho_+\geq\rho_{res}>\rho_-$ & $P_{H1},~P_{H2},~P_s$
                 \\ \cline{2-5} 
  & $\rho_{res}=\rho_b$ & $P_{H1},~P_s$
       & $\rho_-\geq\rho_{res}>\rho_b$ & $P_{H1},~P_{H2},~P_s,~P_{H3}$
                 \\ \cline{2-5} 
  & $\rho_b>\rho_{res}>\rho_-$ & $P_s$
       & $\rho_{res}=\rho_b$ & $P_{H1},~P_s~P_{H3}$ \\ \cline{2-5}
  & $\rho_-\geq\rho_{res}$ & $P_s,~P_{H3}$
     & $\rho_b>\rho_{res}$ & $P_s,~P_{H3}$ \\ \hline
\end{tabular}
\end{center}
\end{table}

Because the Daya Bay experimental result, $\theta_{13}= 8.8^{\circ}\pm0.8^{\circ}$, is quite different from the condition $\theta_{13}< 3^{\circ}$, it is necessary to make clear the behavior of $P_H$ for large $\theta_{13}$. From Eqs. (\ref{PH})-(\ref{gamma}), it can be found that $P_H$ depends on $F$ and $\gamma$ which is related to $\theta_{13}$ and the neutrino energy $E$.

Fig. 3, Fig. 4, and Fig. 5 illustrate the crossing probability $P_H$ as a function of the neutrino energy $E$, the time $t$ and the mixing angle $\theta_{13}$, respectively. In Fig. 3, it can be seen that the value of $P_H$ depends on the energy of SN neutrino. Whatever value $\theta_{13}$ takes, the value of $P_H$ has a great "jump" approximately at $E=10MeV$. For $\theta_{13}=3^{\circ}$, the curve of $P_H$ still has obvious continuous fluctuations from about $15MeV$ to higher energy. For $\theta_{13}=6^{\circ}$ and $\theta_{13}=9^{\circ}$, the value of $P_H$ changes smoothly and decreases slowly when $E\geq30MeV$. Fig. 4 shows the curves of $P_H$ for three typical neutrino energies when the time ranges from $0s$ to $10s$. We can see that as the energy increases the curve becomes fatter. In other words, the greater the neutrino energy, the longer time $P_H$ keeps at high values. In general, the value of $P_H$ reaches the maximum value when the time is between $4-6s$. From Fig. 5(a), it is found that for a certain neutrino energy, at different times, the value of $P_H$ changes smoothly in the range of $\theta_{13}=5^{\circ}-10^{\circ}$. However, the curve for $t=6s$ has rapid fluctuations between $0^{\circ}$ and $5^{\circ}$. In Fig. 5(b), all the curves corresponding to three neutrino energies have obvious fluctuations when $\theta_{13}$ is between $0^{\circ}$ and $5^{\circ}$. This is far from the real $\theta_{13}$ value. It can be seen from Figs. 3, 4, and 5 that $P_H$ is zero near the real value of $\theta_{13}(8.8^{\circ})$ when there are no shock wave effects. However, when the shock wave effects turn on $P_H$ is not zero in a range of time.

\subsection{Collective effects and survival probability $P_{\nu\nu}$}

The neutrino collective effects, which mechanism is totally different from the MSW effects, are caused by the neutrino-neutrino interaction inside the core-collapse SN. Recently, it has been realized to be a crucial feature at very high densities of the core and can change the emitted fluxes of different flavor neutrinos substantially \cite{Dasgupta}\cite{Chakraborty}.

Up to now, there have been a significant amount of studies on the neutrino collective effects \cite{Dasgupta}\cite{Chakraborty}\cite{Fogli0}\cite{Duan1}-\cite{Fogli3}. In Ref. \cite{Chakraborty}, it was shown that the collective effects depend on the inherent features of SN neutrinos, such as their initial total energy, relative luminosities of different flavors, and the neutrino mass hierarchy \cite{Kotake}\cite{Lunardini2}\cite{Takahashi2}.

In order to study the collective effects quantitatively, we set $P_{\nu\nu}$ as the survival probability that the neutrinos (antineutrinos) $\nu(\bar{\nu})$ remain as their original states after the collective effects. In Ref. \cite{Dasgupta} the authors introduced a simplified picture to describe the characteristics of the collective effects:
\begin{equation}
P_{\nu\nu}=
\begin{cases} 1 & (E<E_C), \\
 0 & (E>E_C),
 \end{cases}
\label{Pnunu1}
\end{equation}
for neutrinos and
\begin{equation}
\bar{P}_{\nu\nu}=1. \label{Pnunu2}
\end{equation}
for antineutrinos, where $E_C$ is a critical energy. We take $E_C=10MeV$ in our later calculation \cite{Dasgupta}.

\subsection{Earth matter effects}

In this subsection, we briefly review known analytical results about the Earth matter effects, which were studied in Ref. \cite{Guo} for the sake of completeness and self-consistency of this paper. If a neutrino reaches the detector with an incident angle $\alpha$ (see Fig. 1 in Ref. \cite{Huang}), the distance that the neutrino travels through the Earth to the detector, $L$, and the distance of the neutrino to the center of the earth, $\tilde{x}$, can be given by
\begin{eqnarray}
 L&=&(-R_E+h)\cos{\alpha}+\sqrt{R_E^2-(R_E-h)^2\sin^2{\alpha}},
 \nonumber\\
 \tilde{x}&=&\sqrt{(-R_E+h)^2+(L-x)^2+2(R_E-h)(L-x)\cos{\alpha}},
 \nonumber
\end{eqnarray}
where $h$ is the depth of the detector in the Earth, $x$ the distance that the neutrino travels into the Earth, and $R_E$ the radius of the Earth.

Let $P_{ie}$ be the probability that a neutrino mass eigenstate $\nu_i$ enters the surface of the Earth and arrives at the detector as an electron neutrino $\nu_e$, then we have \cite{Ioannisian}
\begin{equation}
P_{2e}=\sin^2\theta_{12}+\frac{1}{2}\sin^22{\theta_{12}}\int_{x_0}^{x_f}{\rm
d}xV_E(x)\sin\phi
_{x\rightarrow x_f}^{m}, \label{P2e}
\end{equation}
where $V_E(x)$ is the potential $\nu_e$ experiences due to the matter density $\rho_E(x)$ inside the Earth
\cite{Guo}\cite{Dziewonski}\cite{Ioannisian}
\begin{eqnarray}
V_E(x)=\sqrt{2}G_F N_A \rho_E(x) Y_e, \label{VE}
\end{eqnarray}
and $\phi_{a\rightarrow b}^m$ is defined as
\begin{eqnarray}
\phi_{a\rightarrow b}^m &=& \int_a^b {\rm d}x \Delta_m(x),\nonumber
\end{eqnarray}
where
\begin{eqnarray}
\Delta_m(x)&=&\frac{\Delta m_{21}^2}{2E}
 \sqrt{(\cos2\theta_{12}-\varepsilon(x))^2+ \sin^22\theta_{12}},
 \label{detlam}
\end{eqnarray}
with $\theta_{12}\backsimeq34.5^{\circ}$, $\Delta
m^2_{21}=7.7\times10^{-5}eV^2$, $\varepsilon(x)={2EV_E(x)}/{\Delta
m_{21}^2}$ \cite{Gonzalez-Garcia}.
We will use the realistic matter density profile inside the Earth to compute the Earth matter effects.

\section{Detecting SN neutrinos on the Earth}

As mentioned in Refs. \cite{Guo}\cite{Huang}, there are three reaction channels with which one may detect SN neutrinos at the Daya Bay experiment. In this section, we review the three reactions and the calculation methods of event numbers. The reaction formulas and the means of calculation are still applicable when $\theta_{13}$ is large.

\subsection{Three reactions}

A detailed description of the Daya Bay experiment can be found in Refs. \cite{An2}\cite{Guo2}. There are eight detectors located at different sites of the Daya Bay experiment. The total detector mass is about 300 tons and the depth of the detectors $h\backsimeq400$ m. The Daya Bay Collaboration uses Linear Alkyl Benzene (LAB) as the main part of the liquid scintillator. LAB has a chemical composition including $C$ and $H$ with the ratio of the number of $C$ and $H$ being about 0.6. Then, the total numbers of target protons, electrons, and $^{12}C$ are
\begin{eqnarray}
N_T^{(p)}=2.20\times10^{31}, \quad N_T^{(e)}=1.01\times10^{32},
\quad N_T^{(C)}=1.32\times10^{31} \nonumber.
\end{eqnarray}

In the Daya Bay experiment, there are three reactions which can be used to detect SN neutrinos: the inverse beta decay, neutrino-electron reactions, and neutrino-carbon reactions. Their effective cross sections are given as follows \cite{Cadonati}\cite{Burrows}.

(1) The cross section for the inverse beta-decay is
\begin{eqnarray}
 \sigma(\bar{\nu}_ep)=9.5\times10^{-44}(E(MeV)-1.29)^2 cm^2, \label{sigma p}
\end{eqnarray}
where the reaction threshold is $E_{th}=1.80MeV$.

(2) The neutrino-electron scattering interactions can be divided into changed current and neutral current ones. The total cross sections have the following forms:
\begin{eqnarray}
\langle\sigma({\nu_e}e\rightarrow{\nu_e}e)\rangle=9.2\times10^{-45}E(MeV)^2
cm^2,\nonumber\\
\langle\sigma({\bar{\nu}_e}e\rightarrow{\bar{\nu}_e}e)\rangle=3.83\times10^{-45}E(MeV)^2
cm^2,\nonumber\\
\langle\sigma({\nu_{\mu,\tau}}e\rightarrow{\nu_{\mu,\tau}}e)\rangle=1.57\times10^{-45}E(MeV)^2
cm^2,\nonumber\\
\langle\sigma({{\bar{\nu}}_{\mu,\tau}}e\rightarrow{{\bar{\nu}}_{\mu,\tau}}e)\rangle=1.29\times10^{-45}
E(MeV)^2cm^2.
\end{eqnarray}

(3) The cross sections for the neutrino-carbon reactions are
\begin{eqnarray}
 &&\ \langle\sigma(^{12}C(\nu_e,e^{-})^{12}N)\rangle=1.85\times10^{-43}E(MeV)^2
cm^2,\nonumber \\
&&\
\langle\sigma(^{12}C(\bar{\nu}_e,e^{+})^{12}B)\rangle=1.87\times10^{-42}E(MeV)^2
cm^2, \label{sigma CB}
\end{eqnarray}
for the charged-current capture,  and
\begin{eqnarray}
&&\ \langle\sigma(\nu_e^{12}C)\rangle=1.33\times10^{-43}E(MeV)^2cm^2, \nonumber \\
&&\ \langle\sigma(\Bar{\nu}_e^{12}C)\rangle=6.88\times10^{-43}E(MeV)^2cm^2,
\label{sigma CC} \\
&&\
\langle\sigma(\nu_{\mu,\tau}(\Bar{\nu}_{\mu,\tau})^{12}C)\rangle=3.73\times10^{-42}E(MeV)^2cm^2,
\nonumber
\end{eqnarray}
for the neutral-current capture.

We should note that when neutrino oscillations are taken into account, the oscillations of higher energy $\nu_x$ into $\nu_e$ result in an increased event rate since the expected $\nu_e$ energies in the absence of oscillations are just at or below the charged-current reaction threshold. This leads to an increase by a factor of 35 for the cross section $\langle\sigma(^{12}C(\nu_e,e^-)^{12}N)\rangle$ if we average it over a $\nu_e$ distribution with $T=8MeV$ rather than $3.5MeV$. Similarly, the cross section $\langle\sigma(^{12}C(\Bar{\nu}_e,e^+)^{12}B)\rangle$ is increased by a factor of 5.

In the following subsections, we will discuss the SN neutrino fluxes taking into account the four physical effects and calculate event numbers.

\subsection{Calculation of event numbers}

With all of the four physical effects being taken into account, the SN neutrino fluxes at the detector are expressed as
\begin{eqnarray}
F_{\nu_e}^D&=&pF_{\nu_e}^{(0)}+(1-p)F_{\nu_x}^{(0)},
\nonumber \\
F_{\bar{\nu}_e}^D&=&\bar{p}F_{\bar{\nu}_e}^{(0)}+
  (1-\bar{p})F_{\bar{\nu}_x}^{(0)},
\nonumber \\
2F_{\nu_x}^D&=&(1-p)F_{\nu_e}^{(0)}+(1+p)F_{\nu_x}^{(0)},
\nonumber \\
2F_{\bar{\nu}_x}^D &=&(1-\bar{p})F_{\bar{\nu}_e}^{(0)}+
 (1+\bar{p})F_{\bar{\nu}_x}^{(0)},
\label{FD}
\end{eqnarray}
where the survival probabilities $p$ and $\bar{p}$ are given by
\begin{eqnarray}
p&=&P_{2e}[P_HP_{\nu\nu}+(1-P_H)(1-P_{\nu\nu})], \nonumber\\
\bar{p}&=&(1-\bar{P}_{2e})\bar{P}_{\nu\nu}, \label{pn}
\end{eqnarray}
for the normal mass hierarchy and
\begin{eqnarray}
p&=&P_{2e}P_{\nu\nu}, \nonumber\\
\bar{p}&=&(1-\bar{P}_{2e})[\bar{P}_H\bar{P}_{\nu\nu}+
  (1-\bar{P}_H)(1-\bar{P}_{\nu\nu})],
\label{pi}
\end{eqnarray}
for the inverted mass hierarchy.

Therefore, the event numbers $N(i)$ of SN neutrinos in the reaction channel "$i$" can be calculated by integrating over the neutrino energy, the product of the target number $N_T$, the cross section of the given channel $\sigma(i)$, and the neutrino flux function at the detector $F_{\alpha}^D$,
\begin{equation}
 N(i)=N_T\int{{\rm d}E\cdot\sigma(i)\cdot\frac{1}{4\pi
 D^2}\cdot F_{\alpha}^D}, \label{Ntotal}
\end{equation}
where $\alpha$ stands for the neutrino or antineutrino of a given flavor, and $D$ is the distance between the SN and the Earth.

\subsection{Scanning over the relevant parameters}

In this subsection, we scan the relevant parameters in the two parametrization forms for neutrino energy distribution. In Section II, we gave the ranges of the neutrino temperatures and pinching parameters of the spectra in the Fermi-Dirac distribution, as listed in Eq. (\ref{T}), and the ranges of the average energies of neutrinos and dimensionless pinching parameters in the parametrization of SN neutrino fluxes given by the Garching group, as listed in Eq. (\ref{E}). It is expected to obtain the maximum and minimum values of neutrino event numbers in the Daya Bay experiment from our calculation results. To achieve this objective, scanning over the ranges of all parameters related to the calculation of four physical effects on detecting SN neutrinos is necessary. Notice that the luminosity ratios of different flavor neutrinos in Eq. (\ref{L1}) and Eq. (\ref{L2}) should be considered as well. In fact, as will be shown in the next subsection, the luminosity ratios do have effects on neutrino event numbers.

Comparing the three reaction channels in Eqs. (\ref{sigma p})-(\ref{sigma CC}), it can be seen that the cross sections for the neutrino-electron scattering channel are much smaller than the other two reaction channels \cite{Guo}. Hence, we will only consider the inverse beta-decay and the neutrino-carbon reactions in the following analysis. From Eq. (\ref{sigma p}), it can be seen that the inverse beta-decay does not involve any parameters about $\nu_{e}$ since $\nu_{e}$ is not involved in the inverse beta-decay. Based on simulation results the luminosity of $\nu_{e}$ and $\bar{\nu}_{e}$ can be taken to be equal \cite{Keil}, so we can define

\begin{eqnarray}
\frac{L_{{\nu}_{e}}}{L_{{\nu}_{x}}}=  \frac{L_{\bar{\nu}_{e}}}{L_{{\nu}_{x}}}=\frac{1}{M}.
\label{M}
\end{eqnarray}

Scanning over the ranges of all the parameters, we have numerical results for the event numbers. For the Fermi-Dirac distribution in Eq. (\ref{Foa}), we find the following two groups of parameters correspond to the maximum and minimum event numbers, respectively,
\begin{eqnarray}
(Max) ~~T_{\bar{\nu}_{e}}=5MeV, ~T_{{\nu}_{x}}=9MeV,
~\eta_{\bar{\nu}_{e}}=2, ~\eta_{{\nu}_{x}}=2, ~M=2,
\label{P1}\
\end{eqnarray}
\begin{eqnarray}
(Min) ~~T_{\bar{\nu}_{e}}=5MeV, ~T_{{\nu}_{x}}=7MeV,
~\eta_{\bar{\nu}_{e}}=2, ~\eta_{{\nu}_{x}}=0, ~M=0.5,
\label{P2}\
\end{eqnarray}
for the inverse beta-decay, and
\begin{eqnarray}
(Max) ~~T_{{\nu}_{e}}=3MeV, ~T_{\bar{\nu}_{e}}=5MeV, ~T_{{\nu}_{x}}=7MeV,
~\eta_{{\nu}_{e}}=3, ~\eta_{\bar{\nu}_{e}}=2, ~\eta_{{\nu}_{x}}=0, ~M=2,
\label{C1}\
\end{eqnarray}
\begin{eqnarray}
(Min) ~~T_{{\nu}_{e}}=4MeV, ~T_{\bar{\nu}_{e}}=6MeV, ~T_{{\nu}_{x}}=9MeV, ~\eta_{{\nu}_{e}}=5, ~\eta_{\bar{\nu}_{e}}=2.5, ~\eta_{{\nu}_{x}}=2, ~M=0.5,
\label{C2}\
\end{eqnarray}
for the neutrino-carbon reactions.

From Eq. (\ref{L1}) it can be seen that $M$ varies between two extreme values, $0.5$ and $2$. In order to see the influence of the luminosity ratio itself on event numbers, we only change the values of $M$ in Eqs. (\ref{P1})-(\ref{C2}) to the other extreme values, respectively, while keeping other parameters unchanged, then we obtain the following comparison groups of parameters:
\begin{eqnarray}
(Max-C)  ~~T_{\bar{\nu}_{e}}=5MeV, ~T_{{\nu}_{x}}=9MeV,
~\eta_{\bar{\nu}_{e}}=2, ~\eta_{{\nu}_{x}}=2, ~M=0.5,
\label{PC1}\
\end{eqnarray}
\begin{eqnarray}
(Min-C)  ~~T_{\bar{\nu}_{e}}=5MeV, ~T_{{\nu}_{x}}=7MeV,
~\eta_{\bar{\nu}_{e}}=2, ~\eta_{{\nu}_{x}}=0, ~M=2,
\label{PC2}\
\end{eqnarray}
for the inverse beta-decay reactions, and
\begin{eqnarray}
(Max-C) ~~T_{{\nu}_{e}}=3MeV, ~T_{\bar{\nu}_{e}}=5MeV,  ~T_{{\nu}_{x}}=7MeV,
~\eta_{{\nu}_{e}}=3, ~\eta_{\bar{\nu}_{e}}=2, ~\eta_{{\nu}_{x}}=0, ~M=0.5,
\label{CC1}\
\end{eqnarray}
\begin{eqnarray}
(Min-C) ~~T_{{\nu}_{e}}=4MeV,  ~T_{\bar{\nu}_{e}}=6MeV,  ~T_{{\nu}_{x}}=9MeV,
~\eta_{{\nu}_{e}}=5, ~\eta_{\bar{\nu}_{e}}=2.5,  ~\eta_{{\nu}_{x}}=2, ~M=2,
\label{CC2}\
\end{eqnarray}
for the neutrino-carbon reactions.

In the same way, after scanning over the ranges of all the parameters in Eqs. (\ref{E}) and (\ref{L2}) we obtain the following parameter values for the distribution in Eq. (\ref{Fia}):
\begin{eqnarray}
(Max) ~~\langle E_{\bar{\nu}_{e}}\rangle=12MeV, ~\langle E_{{\nu}_{x}}\rangle=18MeV,
~\beta_{\bar{\nu}_{e}}=3.5, ~\beta_{{\nu}_{x}}=3.5,
~M=2,
\label{PE1}\
\end{eqnarray}
\begin{eqnarray}
(Min) ~~\langle E_{\bar{\nu}_{e}}\rangle=12MeV, ~\langle E_{{\nu}_{x}}\rangle=15MeV,
~\beta_{\bar{\nu}_{e}}=3.5, ~\beta_{{\nu}_{x}}=6,
~M=1.25,
\label{PE2}\
\end{eqnarray}
for the inverse beta decay, and
\begin{eqnarray}
(Max) ~~\langle E_{{\nu}_{e}}\rangle=12MeV, ~\langle E_{\bar{\nu}_{e}}\rangle=12MeV, ~\langle E_{{\nu}_{x}}\rangle=15MeV,
~\beta_{{\nu}_{e}}=3.5, ~\beta_{\bar{\nu}_{e}}=3.5, ~~\beta_{{\nu}_{x}}=6, ~~M=2,
\label{CE1}\
\end{eqnarray}
\begin{eqnarray}
(Min) ~~\langle E_{{\nu}_{e}}\rangle=15MeV, ~\langle E_{\bar{\nu}_{e}}\rangle=15MeV, ~\langle E_{{\nu}_{x}}\rangle=18MeV,
~\beta_{{\nu}_{e}}=3.5, ~\beta_{\bar{\nu}_{e}}=3.5,  ~\beta_{{\nu}_{x}}=3.5,
~M=1.25,
\label{CE2}\
\end{eqnarray}
for the neutrino-carbon reactions.

Similar to Eqs. (\ref{PC1})-(\ref{CC2}), we have the comparison groups of parameters for the distribution in Eq. (\ref{Fia}):
\begin{eqnarray}
(Max-C) ~~\langle E_{\bar{\nu}_{e}}\rangle=12MeV, ~\langle E_{{\nu}_{x}}\rangle=18MeV,
~\beta_{\bar{\nu}_{e}}=3.5,  ~\beta_{{\nu}_{x}}=3.5,
~M=1.25,
\label{PEC1}\
\end{eqnarray}
\begin{eqnarray}
(Min-C) ~~\langle E_{\bar{\nu}_{e}}\rangle=12MeV, ~\langle E_{{\nu}_{x}}\rangle=15MeV,
~~\beta_{\bar{\nu}_{e}}=3.5,  ~~\beta_{{\nu}_{x}}=6,
~~M=2,
\label{PEC2}\
\end{eqnarray}
for the inverse beta decay, and
\begin{eqnarray}
(Max-C)~~\langle E_{{\nu}_{e}}\rangle=12MeV, ~\langle E_{\bar{\nu}_{e}}\rangle=12MeV, ~\langle E_{{\nu}_{x}}\rangle=15MeV,
~\beta_{{\nu}_{e}}=3.5, ~\beta_{\bar{\nu}_{e}}=3.5, ~\beta_{{\nu}_{x}}=6, ~M=1.25,
\label{CEC1}\
\end{eqnarray}
\begin{eqnarray}
(Min-C) ~~\langle E_{{\nu}_{e}}\rangle=15MeV, ~\langle E_{\bar{\nu}_{e}}\rangle=15MeV, ~\langle E_{{\nu}_{x}}\rangle=18MeV,
~\beta_{{\nu}_{e}}=3.5, ~\beta_{\bar{\nu}_{e}}=3.5,  ~\beta_{{\nu}_{x}}=3.5,
~M=2,
\label{CEC2}\
\end{eqnarray}
for the neutrino-carbon reactions.

It is noted that no matter how the value of $M$ changes, the total energy of all flavor neutrinos is a constant and the results with parameters in comparison groups are always between the maximum and minimum event numbers.

\subsection{The SN neutrino event numbers under the influence of four physical effects}

In this subsection, we give the numerical results of SN neutrino event numbers detected at the Daya Bay experiment.

Consider a "standard" supernova at a distance $D=10kpc$ from the Earth, which releases total energy $E_B=3\times10^{58}erg$, and take $\tau=3s$ as the decay time of its luminosity \cite{Spergel}-\cite{Loredo2}. The values of relevant parameters have already been given in detail in previous sections. Given the Daya Bay experimental result, we take $\theta_{13}=8.8^{\circ}$ in our calculations.

Firstly, we calculate the neutrino event numbers with the neutrino spectra of the Fermi-Dirac distribution. With the influence of all the four physical effects being taken into account, the neutrino event numbers $N$ are plotted as a function of the incident angle of the neutrino $\alpha$ in Fig. 6 and Fig. 7 for the inverse beta-decay and the neutrino-carbon interactions, respectively. In order to show the influence of the luminosity ratio, we show the results with "$Max~(Min)$" group of parameters and those with "$Max-C~ (Min-C)$" group of parameters in same figures. For example, in Fig. 6(a) "$Max$" represents the results with parameters in Eq. (\ref{P1}) and "$Max-C$" represents those with parameters in Eq. (\ref{PC1}). It is found that event numbers depend on the mass hierarchy and parameter values. The maximum variation of neutrino event numbers appears at $\alpha\sim93^{\circ}$ when $\alpha$ changes due to the Earth matter effects. The variations in the inverse beta-decay are more obvious than those in the neutrino-carbon reactions.

In Table. III and Table. IV, we sum up the neutrino event numbers for the two reaction channels with the Fermi-Dirac parametrization. The numerical results show that the event numbers and the change rates due to the Earth matter effects depend on the parameters $T_{\alpha}$ and $\eta_{\alpha}$. This was observed previously in Ref. \cite{Guo}. It is found in the present work that when the values of $T_{\alpha}$ and $\eta_{\alpha}$ keep unchanged, the luminosity ratio $M$ plays an important role in determining the event numbers. For instance, comparing the results with the "$Max$" and "$Max-C$" groups of parameters for the inverse beta-decay, in the case of normal mass hierarchy, when $M=0.5$ the event numbers at the incipient angle and the change rate are $110.21$ and $4.81\%$, respectively; while when $M=2$ these two numbers are $132.0$ and $14.08\%$, respectively, which are much larger.

Next, we discuss the results with the "Garching" distribution given in Eq. (\ref{Fia}) with the parameter values being listed in Eqs. (\ref{PE1})- (\ref{CEC2}). Similar to Fig. 6 and Fig. 7, Fig. 8 and Fig. 9 show the behavior of event numbers for the inverse beta-decay and the neutrino-carbon interactions, respectively. The summary of event numbers for the two reaction channels with the "Garching" distribution is given in Table. V and Table. VI. From these figures and tables, we also find that the event numbers and the change rates due to the Earth matter effects depend on the parameters $\langle E_{\alpha}\rangle$ and $\beta_{\alpha}$. Furthermore, the luminosity ratio $M$ has an important influence on the event numbers and the change rates while keeping other parameters unchanged. For example, comparing the results with the "$Max$" and "$Max-C$" groups of parameters for the neutrino-carbon interactions, in the case of normal mass hierarchy, when $M=1.25$ the event numbers at the incipient angle and the change rates are $61.49$ and $0.15\%$, respectively; while when $M=2$ these two numbers are $77.4$ and $0.22\%$, respectively, which are much larger.

Comparing the results with two parametrization forms of neutrino energy distribution, we can easily see that for the inverse beta-decay, event numbers with the Fermi-Dirac distribution are larger than those with the "Garching" distribution. However, for the neutrino-carbon interactions, the event numbers with the Fermi-Dirac distribution are smaller than those with the "Garching" distribution. On the other hand, the change rates of event numbers in the two reaction channels due to the Earth matter effects in the case of the Fermi-Dirac distribution are always larger than those in the case of the "Garching" distribution.

Based on the results in Tables III-VI, in Table VII we give neutrino event numbers detected at the Daya Bay experiment when all the uncertainties are taken into account in the cases of normal and inverted mass hierarchies, respectively. We can see that the event numbers range from $63.66\sim 243.51$ and $16.12\sim 94.9$ for the inverse beta-decay and the neutrino-carbon interactions, respectively.

\section{Summery and Discussions}

Given the new experimental result about $\theta_{13}$ from the Daya Bay Collaboration, we deduce the expression of the neutrino conversion probability in the high resonance region inside SN, $P_H$, in the case of large $\theta_{13}$ $(\backsimeq8.8^{\circ})$, by applying the available knowledge for $P_H$. $P_H$ is expressed in the form of hypergeometric function. In the derivation, we take the shock wave effects into account. We give numerical results of $P_H$ as functions of $\theta_{13}$, $t$, and $E$. It is found that $P_H$ is zero near the real value of $\theta_{13}$ when there are no shock wave effects. However, it is not zero in a certain region of time (roughly $3s\sim8s$ depending on neutino energies) when shock wave effects are taken into account.  Our work is different from previous studies which were usually done in the case of small $\theta_{13}$ $(<3^{\circ})$ \cite{Guo}\cite{Huang}.

We consider the influence of all the four physical effects on the detection of SN neutrinos, including the MSW effects, the SN shock wave effects, the neutrino collective effects, and the Earth matter effects.  Scanning over all the relevant parameters in the two different parametrization forms of neutrino energy distribution, we calculate the event numbers for two reaction channels, the inverse beta-decay and the neutrino-carbon reactions, both of which can be detected at the Daya Bay experiment. It is found that the event numbers depend on the parameters $T_{\alpha}$, $\eta_{\alpha}$, $\langle E_{\alpha}\rangle$, $\beta_{\alpha}$, and $L_{\alpha}$, as well as the mass hierarchy. We emphasize that the event numbers depend on the luminosity ratio substantially. We give the range of SN neutrino event numbers detected at the Daya Bay experiment.

Although a lot of effort has been made on identifying the four physical effects on detection of SN neutrinos, there are still a lot of problems which need to be solved. One example is the neutrino collective effects in the case of large $\theta_{13}$. Progress in this direction will affect the detection of SN neutrinos. Now the Daya Bay II experiment is under consideration. Its detector mass will be about $70$ times of the total detector mass of the Daya Bay experiment. This will make it much more possible to detect SN neutrinos in the future.

\section{Acknowledgments}

This work was supported in part by National Natural Science Foundation of China (Project Numbers 10975018, 11175020, 11275025 and 11205185) and the Fundamental Research Funds for the Central Universities in China.


\newpage

\noindent{\large \bf Figure Captions} \\
\vspace{0.2cm}

\noindent Fig. 1 Numerical result of $F$ as a function of $\theta$ for $n=-3$. The solid and dashed curves represent the results when $F$ takes the expression of Eq. (\ref{F21}) and  Eq. (\ref{FT}), respectively.
\vspace{0.2cm}

\noindent Fig. 2 The changes of the density of the supernova under the influence of the shock wave effects \cite{Huang}\cite{Fogli2}.
\vspace{0.2cm}

\noindent Fig. 3 The crossing probability at the high resonance region
$P_H$ as a function of the neutrino energy $E$ for three neutrino mixing angles at $t=6s$. The solid, dashed and dotted curves correspond to $\theta_{13}=3^{\circ}, 6^{\circ}, 9^{\circ}$, respectively.
\vspace{0.2cm}

\noindent Fig. 4  The crossing probability at the high resonance region $P_H$ as a function of time $t$ for three neutrino energies at $\theta_{13}=9^{\circ}$. The solid, dashed and dotted curves correspond to neutrino energy $E=11, 16, 25MeV$, respectively.
\vspace{0.2cm}

\noindent Fig. 5  The crossing probability at the high resonance region $P_H$ as a function of the neutrino mixing angle $\theta_{13}$: (a) for three different times at $E=11MeV$. The solid, dashed and dotted curves correspond to $t=2s, 4s, 6s$, respectively; (b) for three different neutrino energies at $t=6s$, The solid, dashed, and dotted curves correspond to $E=11, 16, 25MeV$, respectively.
\vspace{0.2cm}

\noindent Fig. 6 The event numbers for the inverse beta-decay interaction with the parametrization form in Eq. (\ref{Foa}). $"\alpha"$ is the incident angle of the SN neutrino reaching the detector. "N(I)" represents normal (inverted) mass hierarchy. In Fig. 6(a) "Max" and "Max-C" correspond to parameter values listed in Eq. (\ref{P1}) and Eq. (\ref{PC1}), respectively; in Fig. 6(b) "Min" and "Min-C" correspond to parameter values listed in Eq. (\ref{P2}) and Eq. (\ref{PC2}), respectively.
\vspace{0.2cm}

\noindent Fig. 7 The event numbers for the neutrino-carbon interactions with the parametrization form in Eq. (\ref{Foa}). $"\alpha"$ is the incident angle of the SN neutrino reaching the detector. "N(I)" represents normal (inverted) mass hierarchy. In Fig. 7(a) "Max" and "Max-C" correspond to parameter values listed in Eq. (\ref{C1}) and Eq. (\ref{CC1}), respectively; in Fig. 7(b) "Min" and "Min-C" correspond to parameter values listed in Eq. (\ref{C2}) and Eq. (\ref{CC2}), respectively.
 \vspace{0.2cm}

\noindent Fig. 8 The event numbers for the inverse beta-decay interaction with the parametrization form in Eq. (\ref{Fia}). $"\alpha"$ is the incident angle of the SN neutrino reaching the detector. "N(I)" represents normal (inverted) mass hierarchy. In Fig. 8(a) "Max" and "Max-C" correspond to parameter values listed in Eq. (\ref{PE1}) and Eq. (\ref{PEC1}), respectively; in Fig. 8(b) "Min" and "Min-C" correspond to parameter values listed in Eq. (\ref{PE2}) and Eq. (\ref{PEC2}), respectively.
\vspace{0.2cm}

\noindent Fig. 9 The event numbers for the neutrino-carbon interactions with the parametrization form in Eq. (\ref{Fia}). $"\alpha"$ is the incident angle of the SN neutrino reaching the detector. "N(I)" represents normal (inverted) mass hierarchy. In Fig. 9(a) "Max" and "Max-C" correspond to parameter values listed in Eq. (\ref{CE1}) and Eq. (\ref{CEC1}), respectively; in Fig. 9(b) "Min" and "Min-C" correspond to parameter values listed in Eq. (\ref{CE2}) and Eq. (\ref{CEC2}), respectively.
\vspace{0.2cm}

\newpage

\begin{figure}
\includegraphics[width=0.5\textwidth]{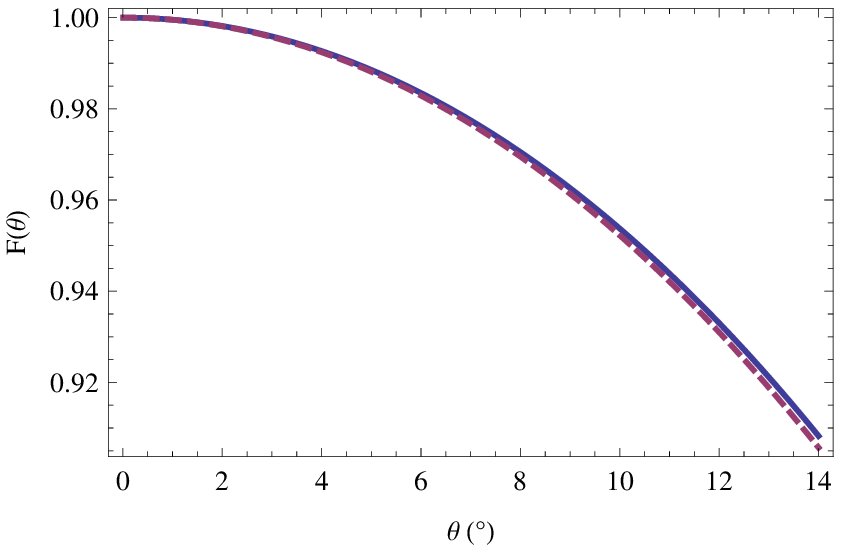}\\
\centerline{Fig. 1}
\end{figure}

\begin{figure}
\includegraphics[width=0.4\textwidth]{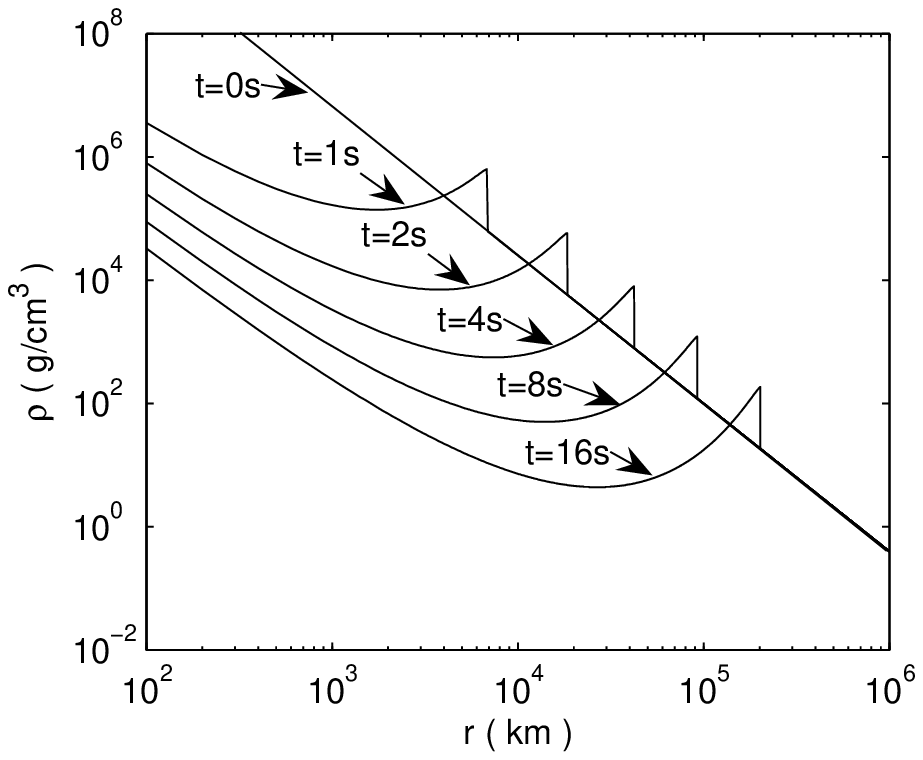}\\
\centerline{Fig. 2}
\end{figure}

\begin{figure}
\includegraphics[width=0.4\textwidth]{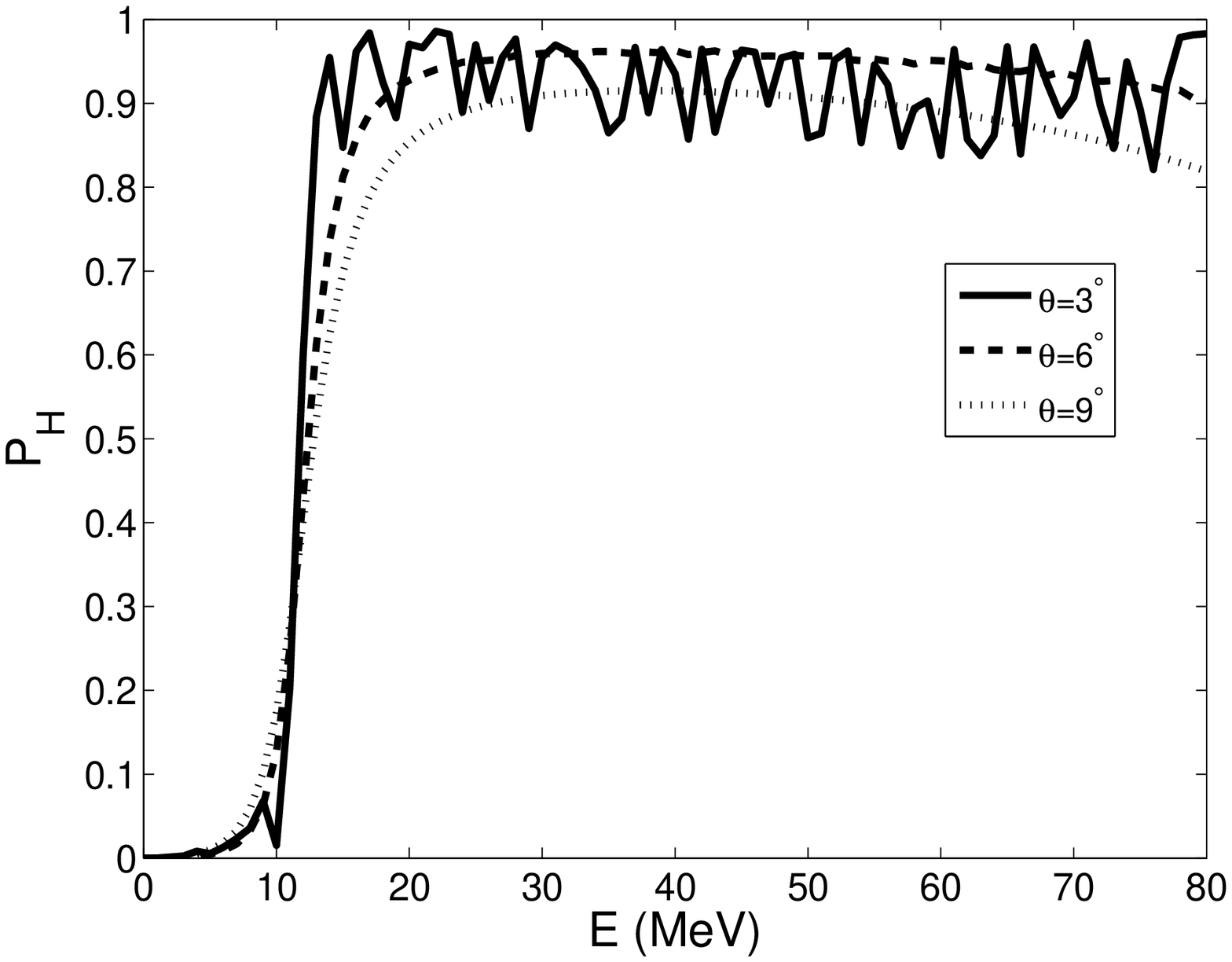}\\
\centerline{Fig. 3}
\end{figure}

\begin{figure}
\includegraphics[width=0.4\textwidth]{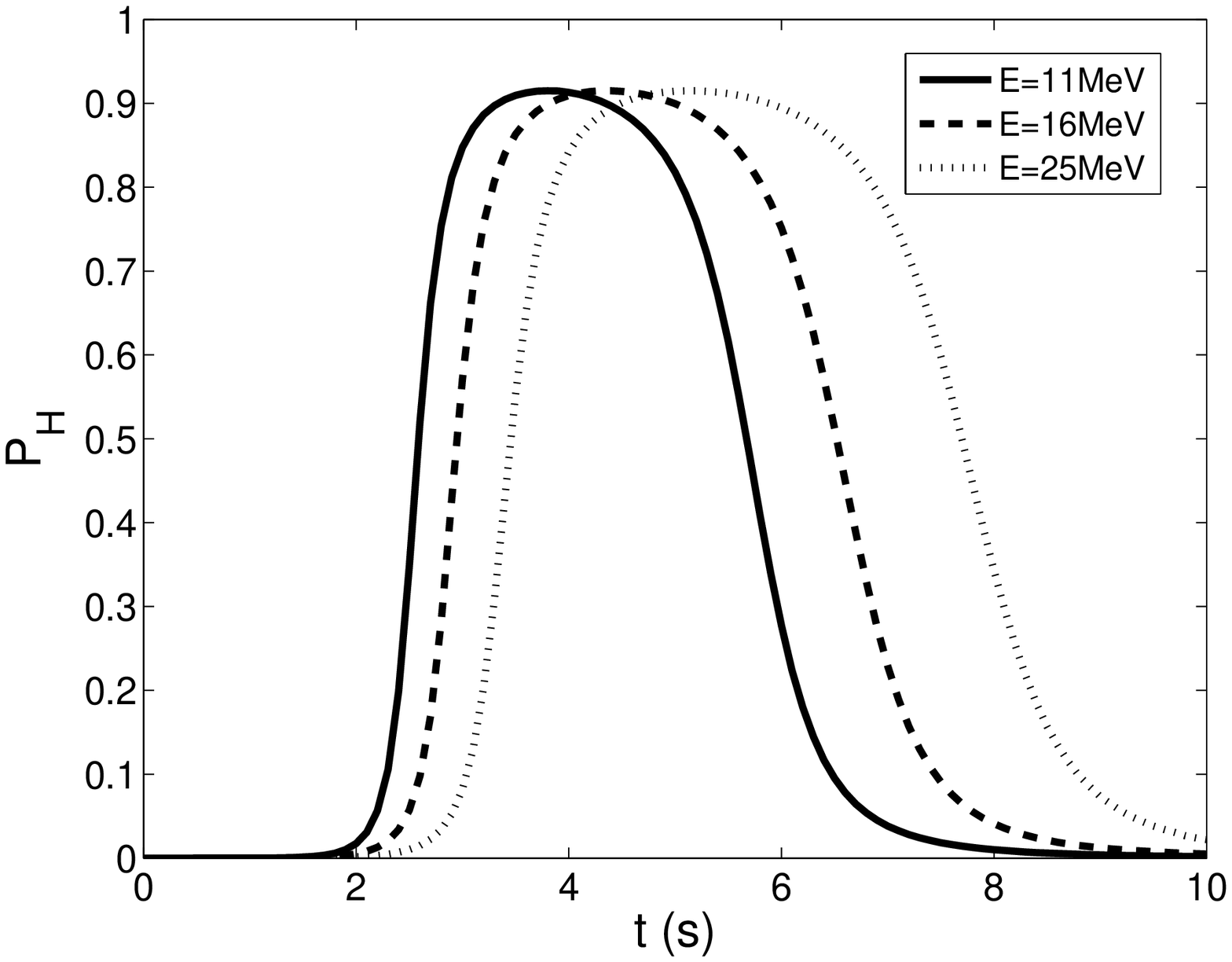}\\
\centerline{Fig. 4}
\end{figure}

\begin{figure}
\begin{minipage}[t]{0.5\linewidth}
\includegraphics[width=0.7\textwidth]{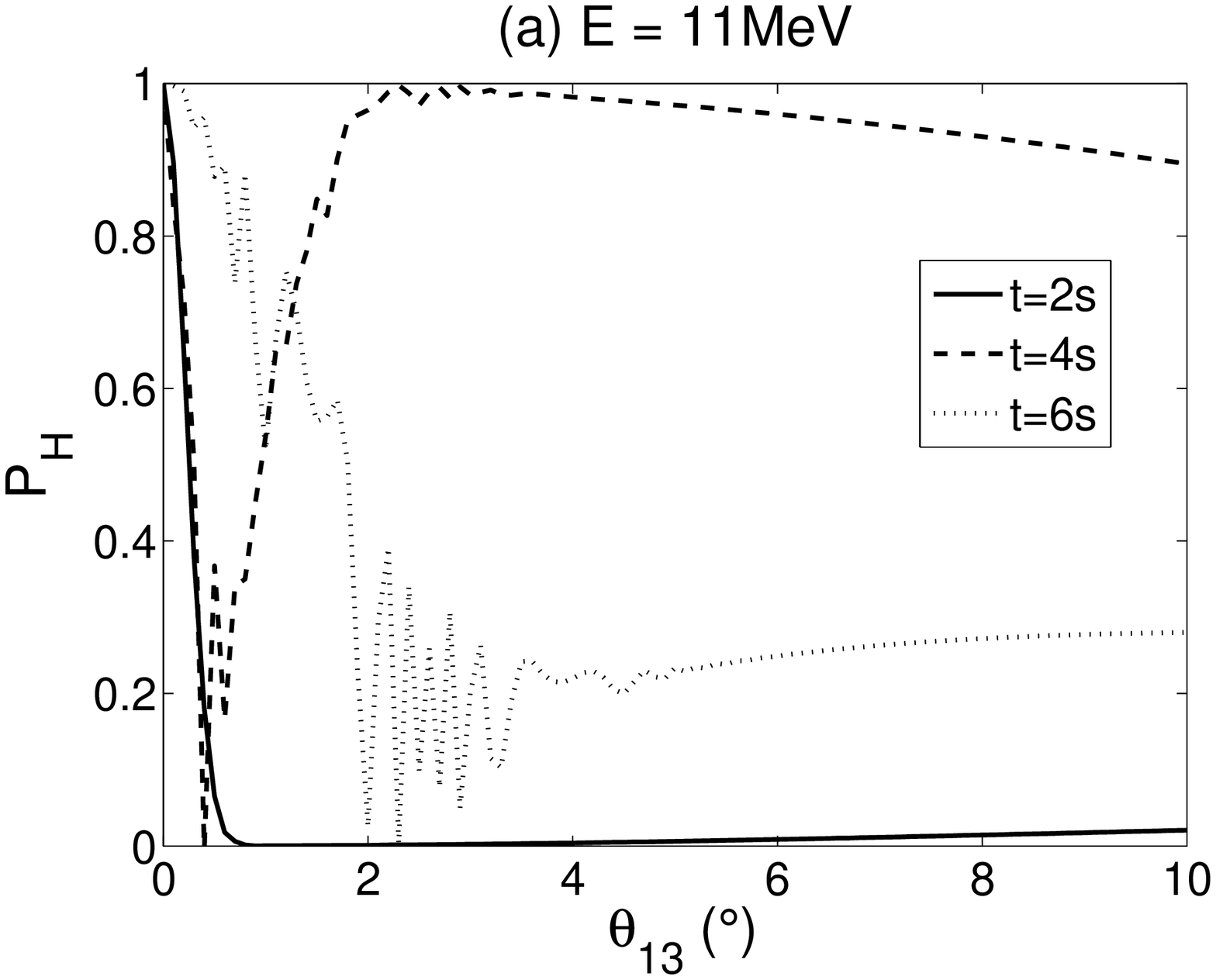}\\
\label{Fig.5(a)}
\centerline{Fig. 5(a)}
\end{minipage}%
\begin{minipage}[t]{0.5\linewidth}
\includegraphics[width=0.7\textwidth]{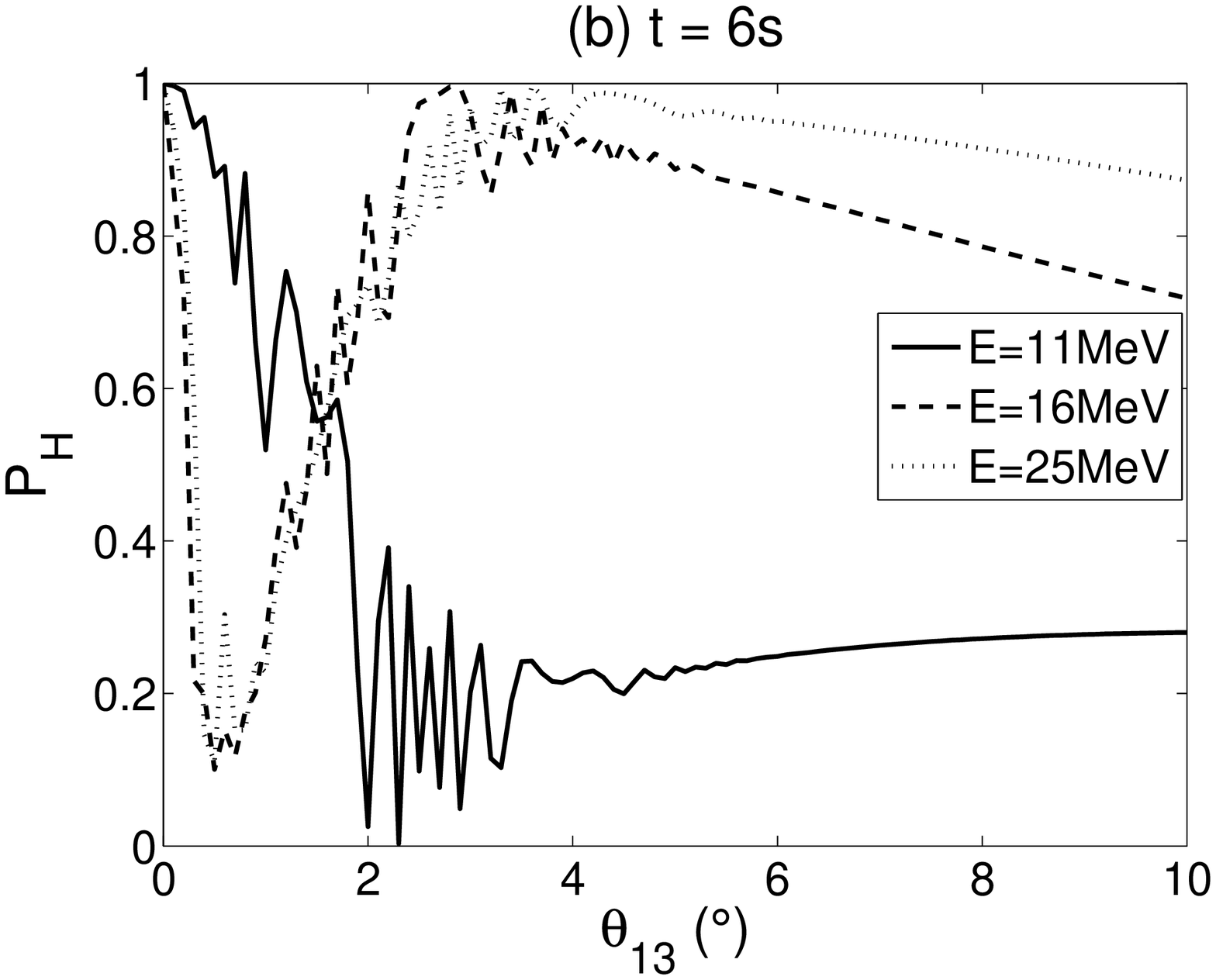}\\
\label{Fig.5(b)}
\centerline{Fig. 5(b)}
\end{minipage}%
\end{figure}

\begin{figure}
\begin{minipage}[t]{0.5\linewidth}
\includegraphics[width=0.59\textwidth]{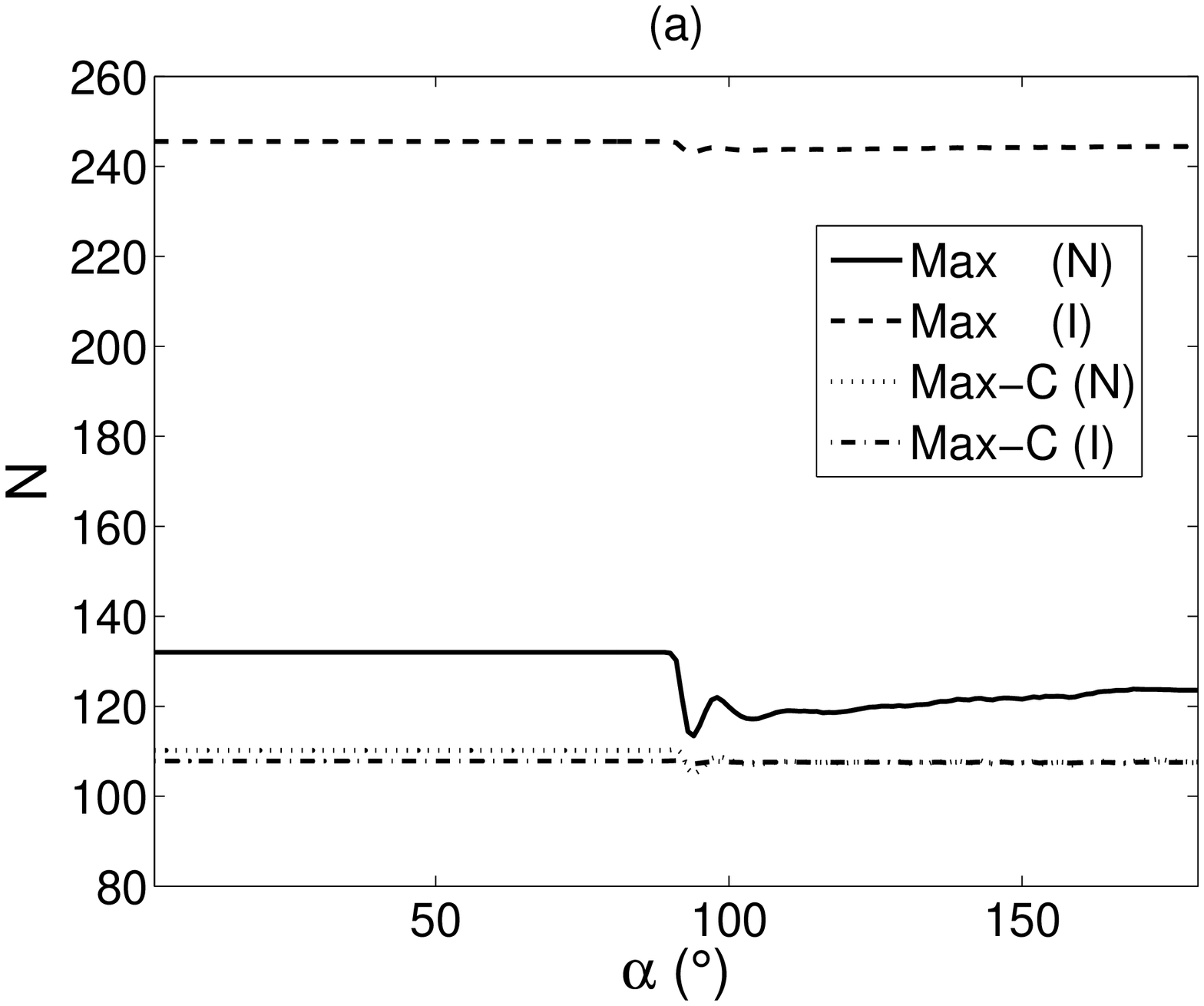}\\
\label{Fig.6(a)}
\centerline{Fig. 6(a)}
\end{minipage}%
\begin{minipage}[t]{0.5\linewidth}
\includegraphics[width=0.59\textwidth]{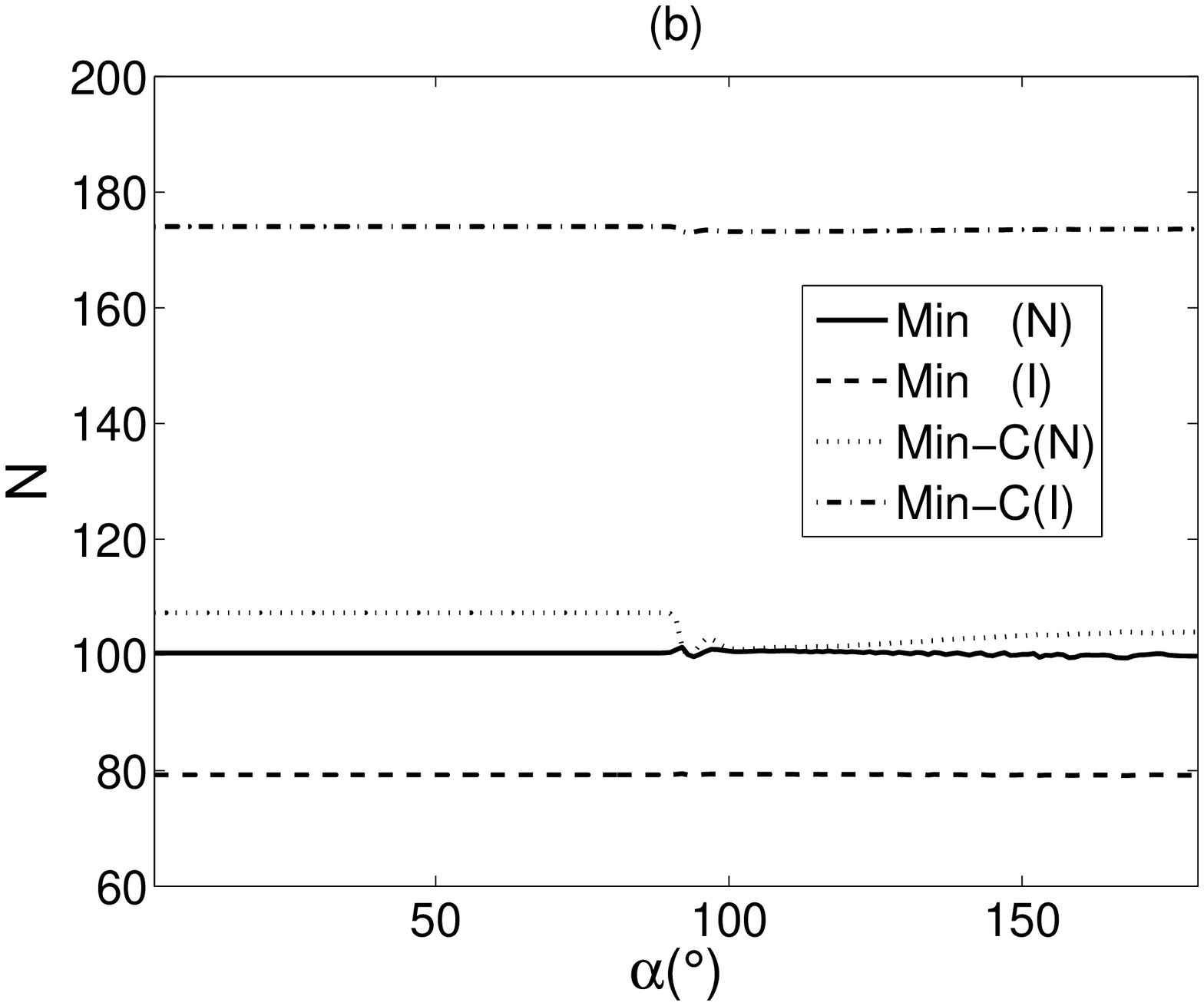}\\
\label{Fig.6(b)}
\centerline{Fig. 6(b)}
\end{minipage}%
\end{figure}

\begin{figure}
\begin{minipage}[t]{0.5\linewidth}
\includegraphics[width=0.59\textwidth]{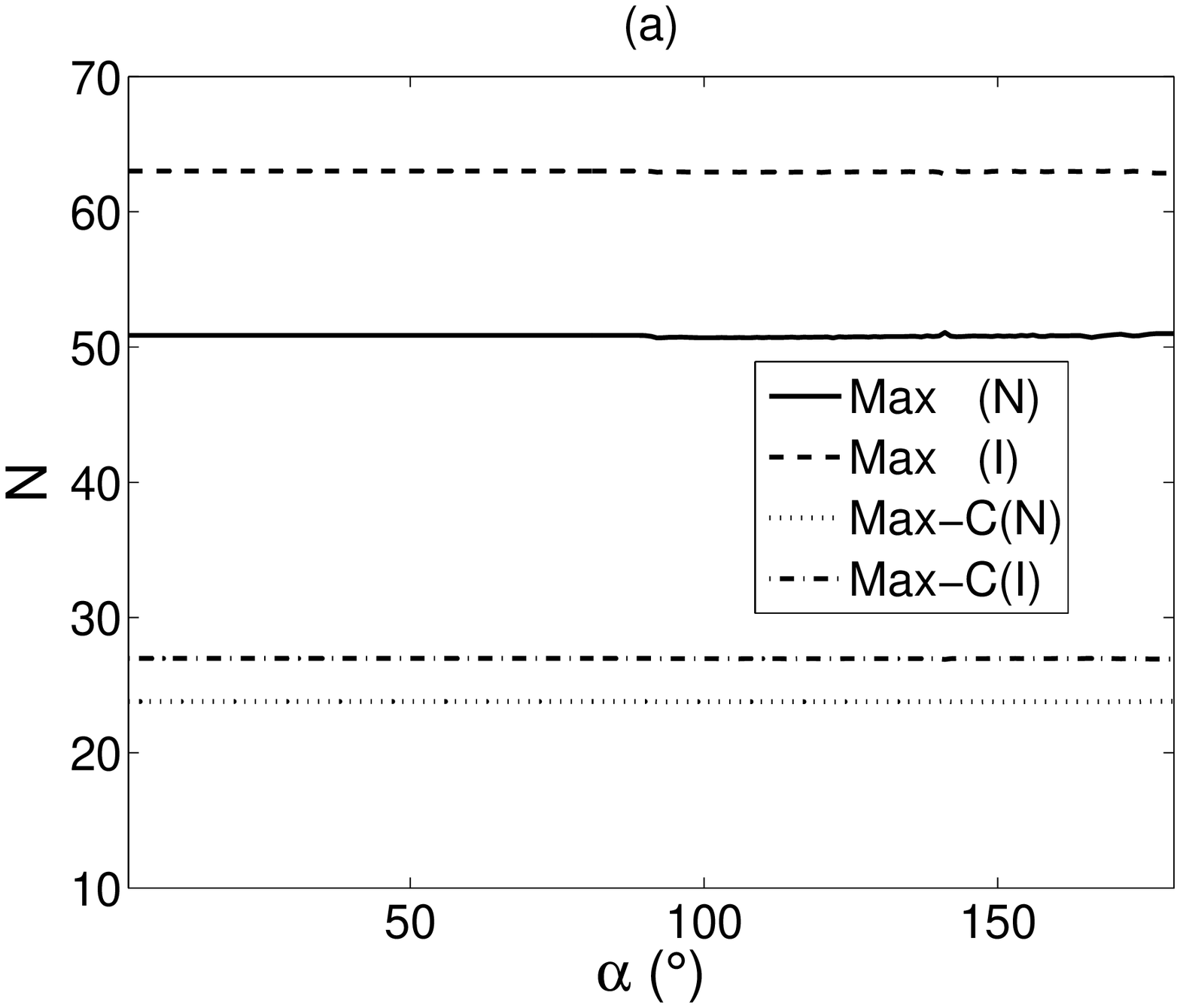}\\
\label{Fig.7(a)}
\centerline{Fig. 7(a)}
\end{minipage}%
\begin{minipage}[t]{0.5\linewidth}
\includegraphics[width=0.59\textwidth]{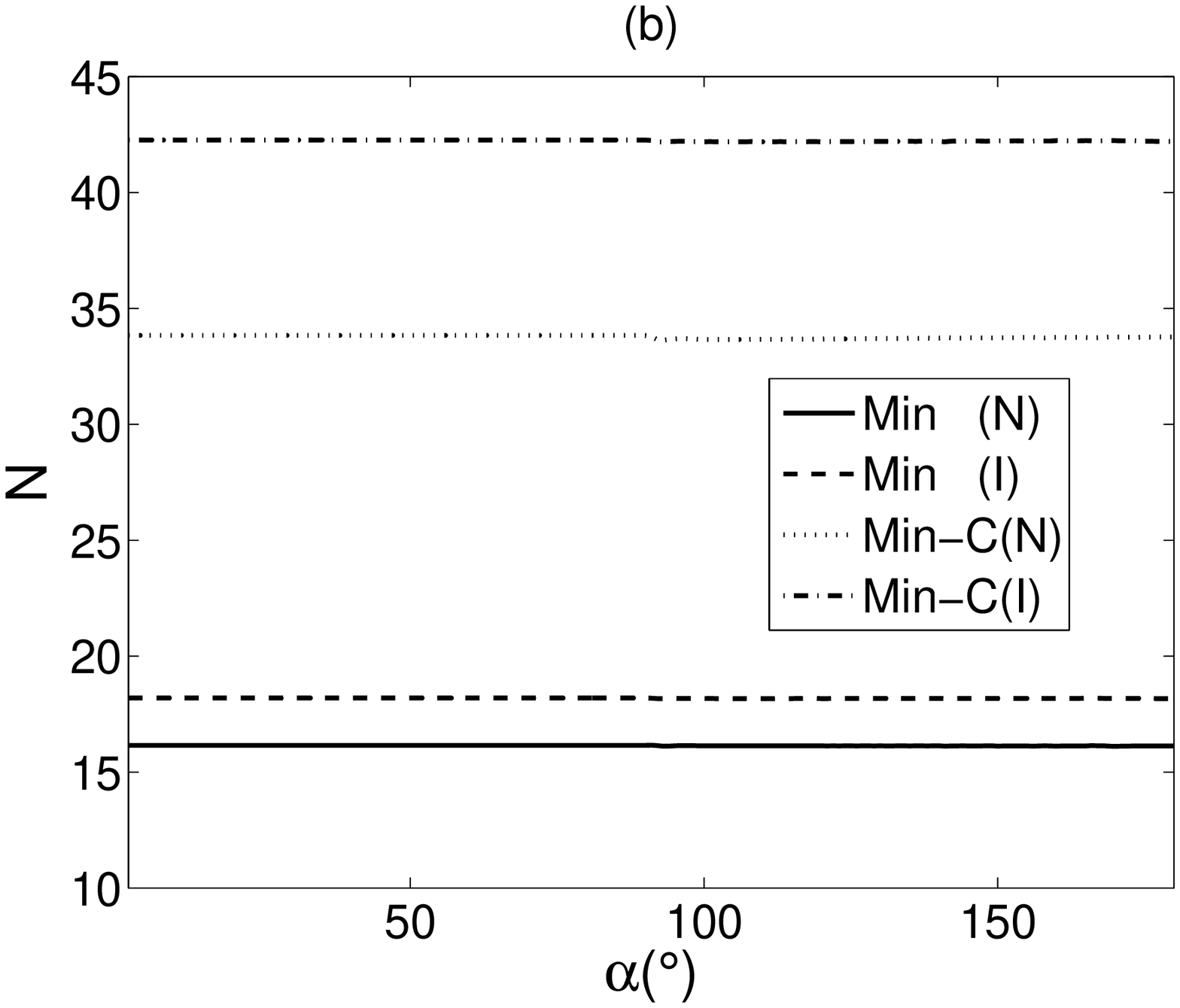}\\
\label{Fig.7(b)}
\centerline{Fig. 7(b)}
\end{minipage}%
\end{figure}

\begin{figure}
\begin{minipage}[t]{0.5\linewidth}
\includegraphics[width=0.59\textwidth]{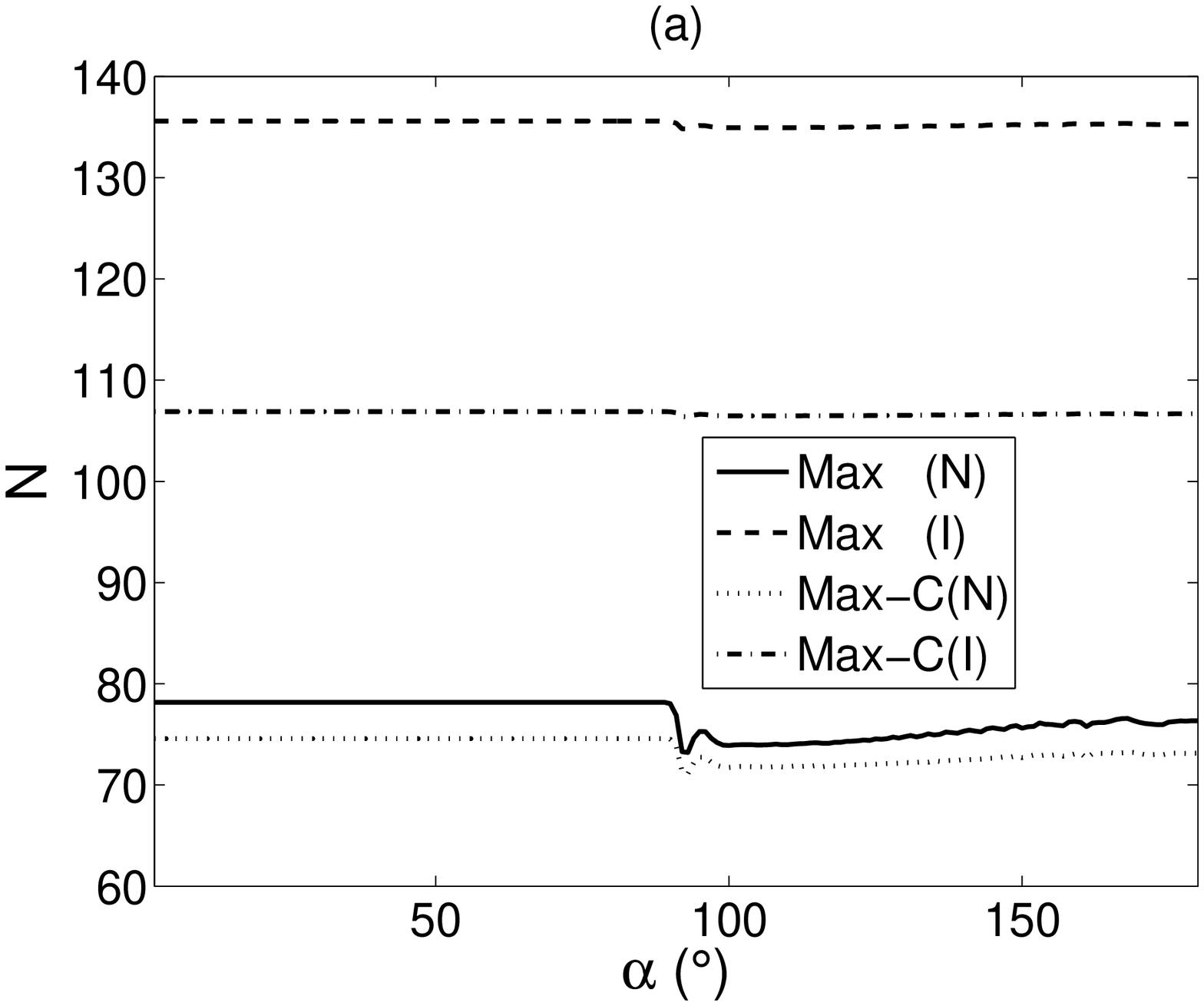}\\
\label{Fig.8(a)}
\centerline{Fig. 8(a)}
\end{minipage}%
\begin{minipage}[t]{0.5\linewidth}
\includegraphics[width=0.59\textwidth]{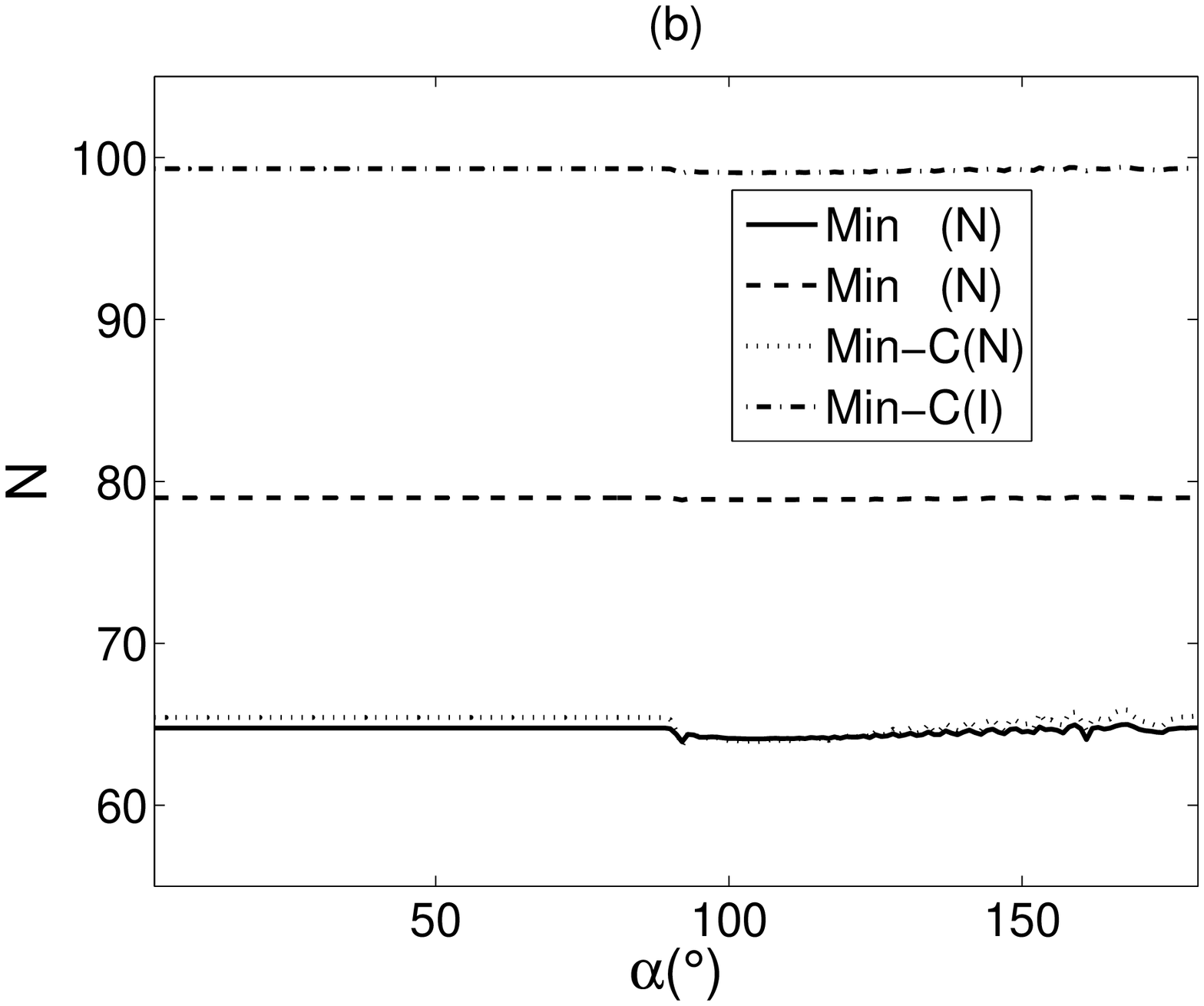}\\
\label{Fig.8(b)}
\centerline{Fig. 8(b)}
\end{minipage}%
\end{figure}

\begin{figure}
\begin{minipage}[t]{0.5\linewidth}
\includegraphics[width=0.59\textwidth]{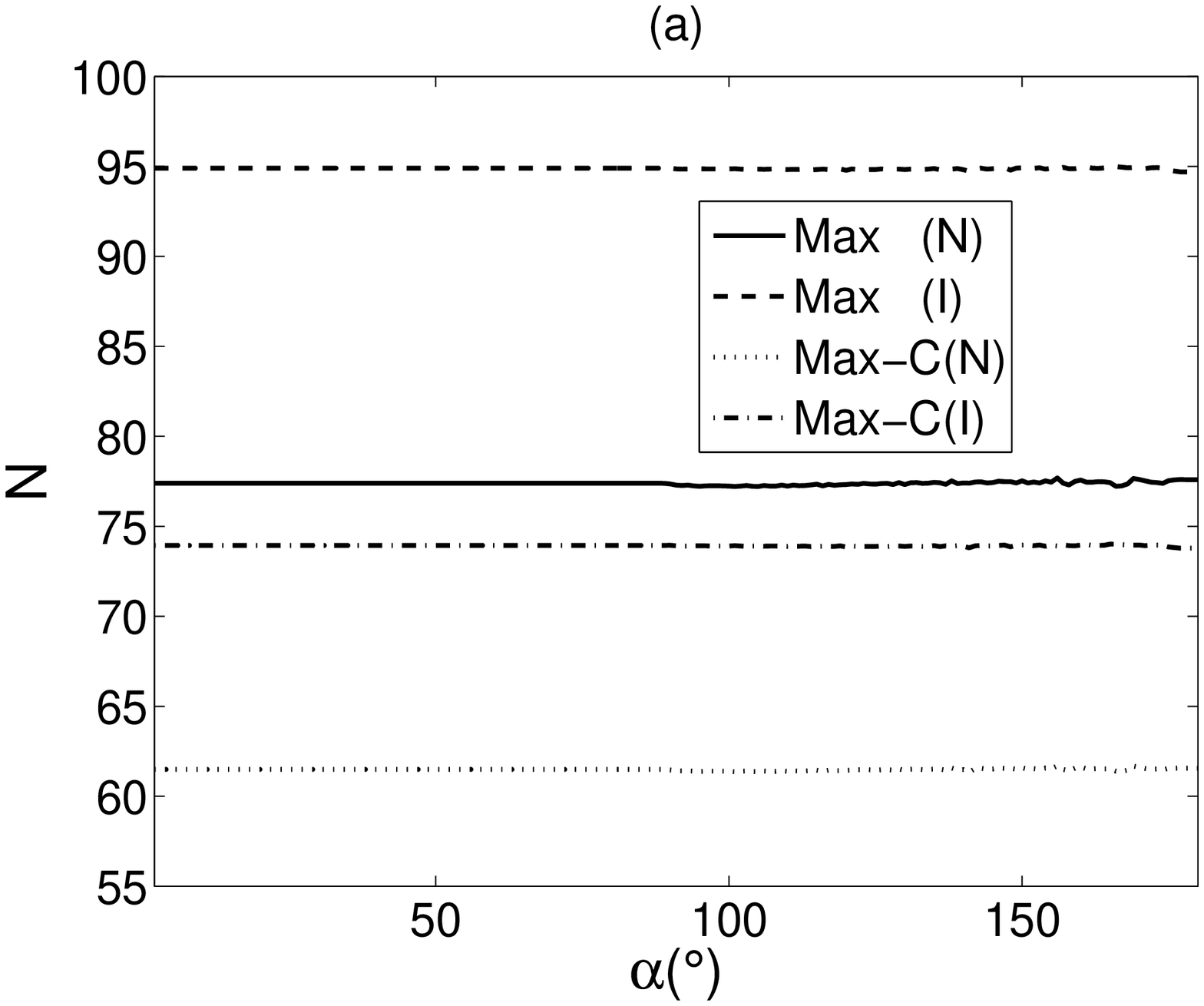}\\
\label{Fig.9(a)}
\centerline{Fig. 9(a)}
\end{minipage}%
\begin{minipage}[t]{0.5\linewidth}
\includegraphics[width=0.59\textwidth]{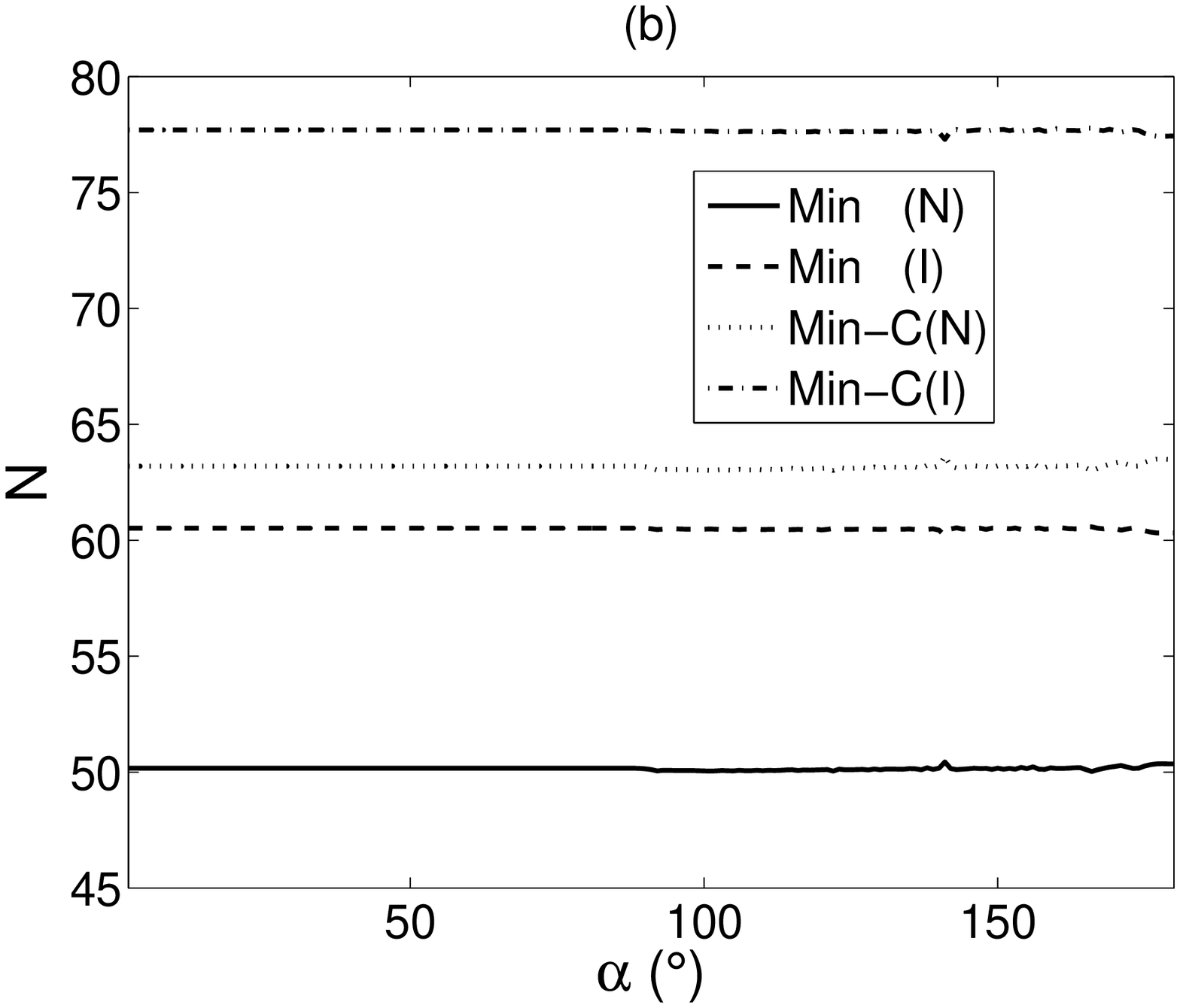}\\
\label{Fig.9(b)}
\centerline{Fig. 9(b)}
\end{minipage}%
\end{figure}

\cleardoublepage

\begin{table}[!htb]
\begin{center}
\caption{Summary of event numbers of the inverse beta-decay interaction for the parametrization form given in Eq. (\ref{Foa}). $"Max"$ and $"Max-C"$ correspond to parameter values in Eq. (\ref{P1}) and Eq. (\ref{PC1}), respectively; $"Min"$ and $"Min-C"$ correspond to parameter values in Eq. (\ref{P2}) and Eq. (\ref{PC2}), respectively. $"N(I)"$ represents normal (inverted) mass hierarchy. The numbers in the columns "Incipient" and "Min" are the event numbers when the SN neutrino incident angle is zero and is the angle in the column "Angle", respectively. The column "Angle" gives the angles at which the event numbers are the minimum and the Earth matter effects are the strongest. "Ratio" gives the percentages of the Earth matter effects.}
\vspace{0.3cm}
\begin{tabular}{|c|c|c|c|c|c|c|}\hline
Reaction & Conditions & Hierarchy & Incipient & Min	& Angle & Ratio \\
\hline
{}  & $Max$  & N & 132.00  &113.41	&94	&14.08\%
 \\
 \cline{3-7}
{} & {} & I & 245.51 &243.13	&94	&0.97\%
 \\
 \cline{2-7}
{} & $Max-C$  & N & 110.21 &104.91	&94	&4.81\%
 \\
 \cline{3-7}
$\bar{\nu}_e+p$ & {}    & I &107.84	&107.22	&94	&0.57\%
 \\
 \cline{2-7}
{}  & $Min$ & N & 100.33	&99.67	&94	&0.66\%
 \\
 \cline{3-7}
{} & {}  & I & 79.26	&79.21	&94	&0.06\%
 \\
 \cline{2-7}
{} & $Min-C$  & N & 107.29 &99.37	&94	&7.38\%
 \\
 \cline{3-7}
{} & {}    & I &174.07 &172.94&94	&0.65\%
 \\
\hline

\end{tabular}
\end{center}
\end{table}

\begin{table}[!htb]
\begin{center}
\caption{Summary of event numbers of the neutrino-carbon interactions for the parametrization form given in Eq. (\ref{Foa}). $"Max"$ and $"Max-C"$ correspond to parameter values in Eq. (\ref{C1}) and Eq. (\ref{CC1}), respectively; $"Min"$ and $"Min-C"$ correspond to parameter values in Eq. (\ref{C2}) and Eq. (\ref{CC2}), respectively. $"N(I)"$ represents normal (inverted) mass hierarchy. The numbers in the columns "Incipient" and "Min" are the event numbers when the SN neutrino incident angle is zero and is the angle in the column "Angle", respectively. The column "Angle" gives the angles at which the event numbers are the minimum and the Earth matter effects are the strongest. "Ratio" gives the percentages of the Earth matter effects.}
\vspace{0.3cm}
\begin{tabular}{|c|c|c|c|c|c|c|}\hline
Reaction & Conditions & Hierarchy & Incipient & Min	& Angle & Ratio \\
\hline
{}  & $Max$  & N & 50.86  &50.68	&92	&0.35\%
 \\
 \cline{3-7}
{} & {} & I & 63.01 &62.92	&92	&0.14\%
 \\
 \cline{2-7}
{} & $Max-C$  & N & 23.79 &23.76	&93	&0.13\%
 \\
 \cline{3-7}
$\nu+^{12}C$& {}    & I &26.97	&26.95	&92	&0.07\%
 \\
 \cline{2-7}
{}  & $Min$ & N & 16.15	&16.11	&93	&0.25\%
 \\
 \cline{3-7}
{} & {}  & I & 18.19	&18.17	&93	&0.11\%
 \\
 \cline{2-7}
{} & $Min-C$  & N & 33.84 &33.62	&93	&0.65\%
 \\
 \cline{3-7}
{} & {}    & I &42.27	 &42.18	&93	&0.21\%
 \\
\hline

\end{tabular}
\end{center}
\end{table}

\begin{table}[!htb]
\begin{center}
\caption{Summary of event numbers of the inverse beta-decay interaction for the parametrization form given in Eq. (\ref{Fia}). $"Max"$ and $"Max-C"$ correspond to parameter values in Eq. (\ref{PE1}) and Eq. (\ref{PEC1}), respectively; $"Min"$ and $"Min-C"$ correspond to parameter values in Eq. (\ref{PE2}) and Eq. (\ref{PEC2}), respectively. $"N(I)"$ represents normal (inverted) mass hierarchy. The numbers in the columns "Incipient" and "Min" are the event numbers when the SN neutrino incident angle is zero and is the angle in the column "Angle", respectively. The column "Angle" gives the angles at which the event numbers are the minimum and the Earth matter effecs are the strongest. "Ratio" gives the percentages of the Earth matter effects.}
\vspace{0.3cm}
\begin{tabular}{|c|c|c|c|c|c|c|}\hline
Reaction & Conditions & Hierarchy & Incipient & Min	& Angle & Ratio \\
\hline
{}  & $Max$  & N & 78.18  &73.21	&93	&6.36\%
 \\
 \cline{3-7}
{} & {} & I &135.58  &134.81	&92	&0.57\%
 \\
 \cline{2-7}
{} & $Max-C$  & N & 74.59 &	71.13 &93	&4.64\%
 \\
 \cline{3-7}
$\bar{\nu}_e+p$ & {}    & I &106.89	&106.38	&93	&0.48\%
 \\
 \cline{2-7}
{}  & $Min$ & N & 64.77	&63.92	&92	&1.31\%
 \\
 \cline{3-7}
{} & {}  & I & 78.98	&78.84	&92	&0.18\%
 \\
 \cline{2-7}
{} & $Min-C$  & N & 65.41 &63.66	&92	&2.68\%
 \\
 \cline{3-7}
{} & {}    & I &99.31 &99.01 &92	&0.30\%
 \\
\hline

\end{tabular}
\end{center}
\end{table}

\begin{table}[!htb]
\begin{center}
\caption{Summary of event numbers of the neutrino-carbon interaction for the parametrization form given in Eq. (\ref{Fia}). $"Max"$ and $"Max-C"$ correspond to parameter values in Eq. (\ref{CE1}) and Eq. (\ref{CEC1}), respectively; $"Min"$ and $"Min-C"$ correspond to parameter values in Eq. (\ref{CE2}) and Eq. (\ref{CEC2}), respectively. $"N(I)"$ represents normal (inverted) mass hierarchy. The numbers in the columns "Incipient" and "Min" are the event numbers when the SN neutrino incident angle is zero and is the angle in the column "Angle", respectively. The column "Angle" gives the angles at which the event numbers are the minimum and the Earth matter effects are the strongest. "Ratio" gives the percentages of the Earth matter effects.}
\vspace{0.3cm}
\begin{tabular}{|c|c|c|c|c|c|c|}\hline
Reaction & Conditions & Hierarchy & Incipient & Min	& Angle & Ratio \\
\hline
{}  & $Max$  & N & 77.40  &77.23	&95	&0.22\%
 \\
 \cline{3-7}
{} & {} & I &94.90  &94.84	&92	&0.06\%
 \\
 \cline{2-7}
{} & $Max-C$  & N & 61.49 &	61.40 &92	&0.15\%
 \\
 \cline{3-7}
$\nu+^{12}C$ & {}    & I &73.94	&73.90	&92	&0.05\%
 \\
 \cline{2-7}
{}  & $Min$ & N & 50.16	&50.05	&92	&0.22\%
 \\
 \cline{3-7}
{} & {}  & I & 60.52	&60.47 &92	&0.08\%
 \\
 \cline{2-7}
{} & $Min-C$  & N & 63.20 &63.03	&92	&0.27\%
 \\
 \cline{3-7}
{} & {}    & I &77.70 &77.62 &92	&0.10\%
 \\
\hline

\end{tabular}
\end{center}
\end{table}

\begin{table}[!htb]
\begin{center}
\caption{The event number ranges in the Daya Bay experiment with all the uncertainties taken into account. $"N(I)"$ represents normal (inverted) mass hierarchy.}
\vspace{0.3cm}
\begin{tabular}{|c|c|c|c|c|}\hline
Reaction & Hierarchy & Max & Min	& Range \\
\hline
$\bar{\nu}_e+p$ & N & 132.00 & 63.66	&{}	
 \\
 \cline{2-4}
{} & I &245.51	&78.84	&$63\sim246$
 \\
\hline
$\nu+^{12}C$ &N & 63.20 &16.11	&{}
 \\
 \cline{2-4}
{} &  I &94.90	 &18.17	&$16\sim95$	
 \\
\hline

\end{tabular}
\end{center}
\end{table}


\end{spacing}

\begin{thebibliography}{}
\bibitem{Kotake} K. Kotake, K. Sato, and K. Takahashi, Rept. Prog. Phys. {\bf 69}, 971 (2006).
\bibitem{Akiyama} S. Akiyama, J. C. Wheeler, D. L. Meier, I. Lichtenstadt, Astrophys. J. {\bf 584}, 954 (2003).
\bibitem{Bethe1} H. A. Bethe, Rev. Mod. Phys. {\bf 62}, 801 (1990);
                 G. E. Brown, H. A. Bethe, and G. Baym, Nucl. Phys. A {\bf 375}, 481 (1982).
\bibitem{Lattimer} J. M. Lattimer, M. Prakash, Astrophys. J. {\bf 550}, 426 (2003).
\bibitem{Bardeen} J. M. Bardeen, B. Carter, S. W. Hawking, Communications in Mathematical Physics. {\bf 31}, 161 (1973).
\bibitem{Lal} D. Lal, Earth and Planetary Science Letters {\bf 104}, 424 (1991).
\bibitem{Arnett1} W. D. Arnett, Astrophys. J. A {\bf 319}, 136 (1987).
\bibitem{Bionta} R. M. Bionta, G. Blewitt, C. B. Bratton, D. Casper and A. Ciocio, {\it et al.}, Phys. Rev. Lett. {\bf 58}, 1494 (1987).
\bibitem{Guo} X.-H. Guo, M.-Y. Huang, and B.-L. Young, Phys. Rev. D {\bf 79}, 113007 (2009).
\bibitem{Huang} M.-Y. Huang, X.-H. Guo, and B.-L. Young, Phys. Rev. D {\bf 82}, 033011 (2010).
\bibitem{Ardellier} F. Ardellier, I. Barabanov, {\it et al.}, arXiv:hep-ex/0606025
\bibitem{Wang} J. M. Wang, {\it et al.}, Astrophy. J. Lett. {\bf 701}, 7 (2009).
\bibitem{Abe1} K. Abe, {\it et al.}, Phys. Rev. Lett. {\bf 107}, 041801 (2011).
\bibitem{Adamson} P. Adamson, {\it et al.}, Phys. Rev. Lett. {\bf 107}, 181802 (2011).
\bibitem{An1} F. P. An, {\it et al.}, Phys. Rev. Lett. {\bf 108}, 171803 (2012).
\bibitem{Abe2} Y. Abe, {\it et al.}, Phys. Rev. Lett. {\bf 108}, 131801 (2012).
\bibitem{Daya} {\it Discovery of new kind of neutrino transformation}, Science Daily,\\
               {\it http://www.sciencedaily.com/releases/2012/03/120308071054.htm}.
\bibitem{Wolfenstein1} L. Wolfenstein, Phys. Rev. D {\bf 17}, 2369 (1978); {\bf 20}, 2634 (1979).
\bibitem{Kuo} T. K. Kuo and J. Pantaleone, Rev. Mod. Phys. {\bf 61}, 937 (1989).
\bibitem{Takiwaki} T. Takiwaki, {\it et al.}, Astrophys. J. {\bf 616}, 1086 (2005).
\bibitem{Fogli1} G. L. Fogli, {\it et al.}, J. Cosmol. Astropart. Phys. {\bf 0504}, 002 (2005).
\bibitem{Dasgupta} B. Dasgupta and A. Dighe, Phys. Rev. D. {\bf 77}, 113002 (2008).
\bibitem{Lunardini1} C. Lunardini and A. Y. Smirnov, Nucl. Phys. B {\bf 616}, 307 (2001).
\bibitem{Mikheyev} S. P. Mikheyev and A. Y. Smirnov, Nucl. Phys. B {\bf 42}, 913 (1985).
\bibitem{Kachelriess} M. Kachelriess, A. Strumia, R. Tomas, J. W. F. Valle, Phys. Rev. D. {\bf 65}, 073016 (2002).
\bibitem{Dziewonski} A. M. Dziewonski and D. L. Anderson, Phys. Earth. Planet. Inter. {\bf 25}, 297 (1981).
\bibitem{Stacey} F. D. Stacey, {\it et al.}, Physics of the Earth, Wiley, New York, (1977).
\bibitem{Chakraborty} S. Chakraborty and S. Choubey, {\it et al.}, J. Cosmol. Astropart. Phys. {\bf 06}, 07 (2010).

\bibitem{Loredo1} T. J. Loredo and D. Q. Lamb, Phys. Rev. D {\bf 65}, 063002 (2002).
\bibitem{Spergel} D. N. Spergel, {\it et al.}, Science {\bf 237}, 1471 (1987).
\bibitem{Fogli0} G. L. Fogli, {\it et al.}, J. Cosmol. Astropart. Phys. {\bf 004}, 030 (2009).
\bibitem{Loredo2} T. J. Loredo and D. Q. Lamb, {\it et al.}, Acad. Sci. {\bf 571}, 601 (1989); T. J. Loredo and D. Q. Lamb, Phys. Rev. D {\bf 65}, 063002 (2002).
\bibitem{Totsuka} Y. Totsuka, Rep. Prog. Phys. {\bf 55}, 377 (1992).

\bibitem{Lunardini2} C. Lunardini and A. Y. Smirnov, J. Cosmol. Astropart. Phys. {\bf 0306}, 009 (2003).
\bibitem{Janka} H.-T. Janka and W. Hillebrandt, Astron. Astrophys. {\bf 224}, 49 (1989);
                Astron. Astrophys. Suppl. Ser. {\bf 78}, 375 (1989);
                H.-T. Janka, Astron. Astrophys. {\bf 244}, 378 (1991).
\bibitem{Keil} M. T. Keil and T. U. Munchen, Preprint astro-ph/0308228;
               M. T. Keil, {\it et al.}, Astrophys. J. {\bf 590}, 971 (2003).
\bibitem{Prudnikov} A. P. Prudnikov, Yu. A. Brychokov and O. I. Marichev, Integrals and Series (Gordon and Breach, New York, 1990).
\bibitem{Schirato} R. C. Schirato and G. M. Fuller, Preprint astro-ph/0205390 (2002).
\bibitem{Takahashi1} K. Takahashi, {\it et al.}, Astropart. Phys. {\bf 20}, 189 (2003).
\bibitem{Tomas} R. Tomas, {\it et al.}, Astropart. Phys. {\bf 0409}, 015 (2004).
\bibitem{Fogli2} G. L. Fogli and E. Lisi, A. Mirizzi and D. Montanino,  Phys. Rev. D {\bf 68}, 033005 (2003).
\bibitem{Gonzalez-Garcia} M. C. Gonzalez-Garcia and M. Maltoni, Phys. Rep. {\bf 460}, 1 (2008); T. Schwetz, M. Tortola, and J. W. F. Valle, New. J. Phys. {\bf 10}, 113011 (2008).

\bibitem{Duan1} H. Y. Duan and J. P. Kneller, J. Phys. G {\bf 36}, 113201 (2009).
\bibitem{Duan2} H. Y. Duan, G. M. Fuller and Y. Z. Qian, Phys. Rev. D {\bf 74}, 123004 (2006); {\bf 76}, 085013 (2007); {\bf 75}, 125005 (2007); Phys. Rev. Lett. {\bf 99}, 241802 (2007).
\bibitem{Hannestad} S. Hannestad, G. G. Raffelt, G. Sigl and Y. Y. Y. Wong, Phys. Rev. D {\bf 74}, 105010 (2006); {\bf 76}, 029901(E) (2007).
\bibitem{Raffelt} G. G. Raffelt and A. Y. Smirnov, Phys. Rev. D {\bf 76}, 081301 (R) (2007); {\bf 76}, 125008 (2007).
\bibitem{Fogli3}G. L. Fogli and E. Lisi, {\it et al.}, J. Cosmol. Astropart. Phys. {\bf 12}, 010 (2007); {\bf 10}, 002 (2009).
\bibitem{Takahashi2} K. Takahashi and K. Sato, Prog. Theor. Phys. {\bf 109}, 919 (2003).

\bibitem{Ioannisian} A. N. Ioannisian and A. Yu. Smirnov, Phys. Rev. Lett. {\bf 93}, 241801 (2004); A. N. Ioannisian, N. A. Kazarian, A. Yu. Smirnov, and D. Wyler, Phys. Rev. D {\bf 71}, 033006 (2005).


\bibitem{An2} F. P. An, {\it et al.}, Nucl. Instrum. Meth. A {\bf 685} 1 (2012).
\bibitem{Guo2} X. H. Guo, {\it et al.} (Daya Bay Collaboration), arXiv:hep-ex/0701029
\bibitem{Cadonati} L. Cadonati, F. P. Calaprice, and M. C. Chen, Astropart. Phys. {\bf 16}, 361 (2002).
\bibitem{Burrows} A. Burrows, S. Reddy, and T. A. Thompson, Nucl. Phys. A {\bf 777}, 356 (2006).




\end{thebibliography}
\end{document}